\documentclass[aps,preprint]{revtex4}%
\usepackage{amsfonts}
\usepackage{amsmath}
\usepackage{amssymb}
\usepackage{graphicx}%
\begin{document}
\preprint{ }
\title[Nuclear Lattice Simulations]{Nuclear Lattice Simulations with Chiral Effective Field Theory}
\author{Dean Lee$^{1}$, Bu\={g}ra Borasoy$^{2}$, and Thomas Schaefer$^{1,3}$}

\affiliation{$^1$Physics Department, North Carolina State University, Raleigh, NC, 27695, USA \\
$^2$Physik Department, Technische Universit\"{a}t M\"{u}nchen D-85747, Garching, Germany \\
$^3$RIKEN-BNL Research Center, Brookhaven National Laboratory, Upton, NY 11973, USA}


\begin{abstract}
We study nuclear and neutron matter by combining chiral effective field theory
with non-perturbative lattice methods. \ In our approach nucleons and pions
are treated as point particles on a lattice. \ This allows us to probe larger
volumes, lower temperatures, and greater nuclear densities than in lattice
QCD. \ The low energy interactions of these particles are governed by chiral
effective theory and operator coefficients are determined by fitting to zero
temperature few-body scattering data. \ Any dependence on the lattice spacing
can be understood from the renormalization group and absorbed by renormalizing
operator coefficients. \ In this way we have a realistic simulation of
many-body nuclear phenomena with no free parameters, a systematic expansion,
and a clear theoretical connection to QCD. \ We present results for hot
neutron matter at temperatures 20 to 40 MeV and densities below twice nuclear
matter density.

\end{abstract}
\maketitle
\tableofcontents

\section{Introduction}

The nuclear many-body problem has long been recognized as one of the central
questions in nuclear physics \cite{Bethe:1971}. The traditional approach to
the many-body problem is based on the assumption that nucleons can be treated
as non-relativistic point-particles interacting mainly via two-body
potentials. Three-body potentials, relativistic effects, and non-nucleonic
degrees of freedom are assumed to give small corrections. The many-body
problem is studied by solving the many-body Schr\"odinger equation. The
ground-state properties of light nuclei and neutron drops have been analyzed
by several groups using variational methods and Green's function Monte Carlo
\cite{Carlson:2003wm}\cite{Wiringa:2002ja}\cite{Pieper:2002ne}%
\cite{Carlson:qn}\cite{Pudliner:1997ck}\cite{Pudliner:1995jn}.

These methods have been very successful, but there are several good reasons
for seeking an alternative approach. One reason is the desire for a theory
that is more directly grounded in QCD. We expect this theory to explain why
two-body forces are dominant, and how the interaction should be chosen. In
addition to that, we would like to have a framework which allows the
calculations to be systematically improved, and provides an estimate of the
errors due to contributions that have been neglected. If we consider the
interaction of a single nucleon with pions and external fields such a
framework is provided by chiral perturbation theory. In a very influential
paper Weinberg proposed to extend effective field theory methods to the
nucleon-nucleon interaction \cite{Weinberg:1990rz}. Over the last several
years effective field theory methods have been applied successfully to the two
and three-nucleon system \cite{Beane:2000fx,Bedaque:2002mn}. Effective field
theory methods have also been applied to nuclear and neutron matter, but these
calculations rely on a perturbative expansion in powers of the Fermi momentum
\cite{Kaiser:2001jx}\cite{Lutz:1999vc}.

Our aim in this work and the goal of the Nuclear Lattice Collaboration as a
whole \cite{Seki:1998qw} is to extend effective field theory methods to the
nuclear many-body problem. For this purpose we investigate the many-body
physics of low-energy nucleons and pions on the lattice. Our starting point is
the same as that of Weinberg.\ We begin with the most general local Lagrangian
involving pions and low-energy nucleons consistent with translational
invariance, isospin symmetry, and spontaneously broken chiral symmetry. This
yields an infinite set of possible interaction terms with increasing numbers
of derivatives and/or nucleon fields. \ Degrees of freedom associated with
anti-nucleons, heavier mesons such as the $\rho$, and heavier baryons such as
the $\Delta$, are integrated out. \ The contribution of these particles appear
as coefficients of local terms in our pion-nucleon Lagrangian. \ We also
integrate out nucleons with momenta greater than $\pi a^{-1}$, where $a$ is
the lattice spacing.

The operator coefficients in our effective Lagrangian are determined by
fitting to experimentally-measured few-body nucleon scattering data at zero
temperature. \ The dependence on the lattice spacing is described by the
renormalization group and can absorbed by renormalizing operator coefficients.
\ In this way we construct a realistic simulation of many-body nuclear
phenomena with no free parameters. \ In our discussion we present results for
hot neutron matter at temperatures 20 to 40 MeV and densities below twice
nuclear matter density.

The first lattice study of nuclear matter was done by Brockmann and
Frank\ \cite{Brockmann:1992in}. \ They used a momentum lattice and analyzed
the quantum hadrodynamics model of Walecka \cite{Walecka:1974qa}.
\ M\"{u}ller, Koonin, Seki, and van Kolck\textit{ }\cite{Muller:1999cp}, were
the first to look at infinite nuclear and neutron matter on a spatial lattice
at finite density and temperature. \ They used an effective nucleon-nucleon
interaction on a $4^{3}$ lattice and found evidence for saturation in nuclear matter.
\ The approach we pursue is similar in spirit to that of \cite{Muller:1999cp}.
\ The main difference is the inclusion of pion degrees of freedom and our use
of chiral effective field theory with Weinberg power counting. \ The nuclear
liquid-gas transition has also been studied using classical lattice gas models
\cite{Qian:2002kj}\cite{Kuo:1996qt}\cite{Ray:1997nn}\cite{Pan:1998ua}.

\section{Notation}

Before describing the physics it will be helpful to first define the notation
that we use throughout our discussion. \ We let $\vec{n}$ represent
integer-valued lattice vectors on our $3+1$ dimensional space-time lattice.
$\ $We use a subscripted `$s$' such as in $\vec{n}_{s}$ to represent purely
spatial lattice vectors. \ We use subscripted indices such as $i,j$ for the
two spin components of the neutron, $\uparrow$ and $\downarrow$. \ We let
$\hat{0}$ be the unit lattice vector in the time direction and let $\hat
{l}_{s}=\hat{1}$, $\hat{2}$, $\hat{3}$ be the corresponding unit lattice
vectors in the spatial directions. \ A summation symbol such as
\begin{equation}
\sum_{l_{s}}%
\end{equation}
implies a summation over values $l_{s}=1$, $2$, $3$.

We let $a$ be the lattice spacing in the spatial direction and $L$ be the
length of the spatial lattice in each direction. \ $a_{t}$ is the lattice
spacing in the temporal direction and $L_{t}$ is the length in the temporal
direction. \ We let $\alpha_{t}$ be the ratio between lattice spacings,%
\begin{equation}
\alpha_{t}=\tfrac{a_{t}}{a}.
\end{equation}
Throughout we use dimensionless parameters and operators, which correspond
with physical values multiplied by the appropriate power of $a$. \ We use the
superscript `$phys$' such as in $m_{N}^{phys}$ to represent quantities with
physical units. \ We use $a,a^{\dagger}$ to represent annihilation and
creation operators for the neutron, whereas $c,c^{\ast}$ indicate the
corresponding Grassmann variables in the path integral representation. \ We
use the symbol $:f(a^{\dagger},a):$ to indicate the normal ordering of
operators in $f(a^{\dagger},a)$. \ We let $m_{N}$ be the mass of the neutron,
$\mu$ be the neutron chemical potential, and $m_{\pi}$ be the mass of the pion.

Our conventions for Fourier transforms are%
\begin{align}
\tilde{f}(\vec{k})  &  =\tfrac{1}{\sqrt{L_{t}L^{3}}}\sum_{\vec{n}}e^{i\vec
{k}_{\ast}\cdot\vec{n}}f(\vec{n}),\\
f(\vec{n})  &  =\tfrac{1}{\sqrt{L_{t}L^{3}}}\sum_{\vec{k}}e^{-i\vec{k}_{\ast
}\cdot\vec{n}}\tilde{f}(\vec{k}),
\end{align}
where%
\begin{equation}
\vec{k}_{\ast}=(\tfrac{2\pi}{L_{t}}k_{0},\tfrac{2\pi}{L}k_{1},\tfrac{2\pi}%
{L}k_{2},\tfrac{2\pi}{L}k_{3}).
\end{equation}
We use periodic boundary conditions in the spatial directions and
periodic/antiperiodic boundary conditions in the temporal direction for bosons/fermions.

We let $D_{N}(\vec{k})\delta_{ij}$ and $D_{\pi}(\vec{k})$ be the free neutron
and neutral pion propagators. \ For notational convenience the spin-conserving
$\delta_{ij}$ in the neutron propagator will from here on be implicit. \ The
self-energies, $\Sigma_{N}(\vec{k})$ and $\Sigma_{\pi}(\vec{k})$, are defined
by%
\begin{equation}
D_{N}^{full}(\vec{k})=\frac{D_{N}(\vec{k})}{1-\Sigma_{N}(\vec{k})D_{N}(\vec
{k})},
\end{equation}%
\begin{equation}
D_{\pi}^{full}(\vec{k})=\frac{D_{\pi}(\vec{k})}{1-\Sigma_{\pi}(\vec{k})D_{\pi
}(\vec{k})},
\end{equation}
where $D_{N}^{full}(\vec{k})$ and $D_{\pi}^{full}(\vec{k})$ are\ the
fully-interacting propagators.

\section{Non-perturbative effective field theory}

Effective field theory provides a systematic method to compute physical
observables order by order in the small parameter $Q/\Lambda_{\chi}$, where
$\Lambda_{\chi}$ is the chiral symmetry breaking scale and $Q=(q,m_{\pi
},\ldots)$. Here, $q$ is a small external momentum and $m_{\pi}$ is the mass
of the pion. The simplest processes are those that involve only pions and
external fields. In this case the effective field theory is perturbative. At
any order in $Q$ there are only a finite number of diagrams that have to be
included. At lowest order these are tree diagrams with the leading order
interaction. At higher order, diagrams with more loops or higher order terms
in the interaction have to be taken into account.

\begin{figure}
[ptb]
\begin{center}
\includegraphics[
height=1.0888in,
width=4.0413in
]%
{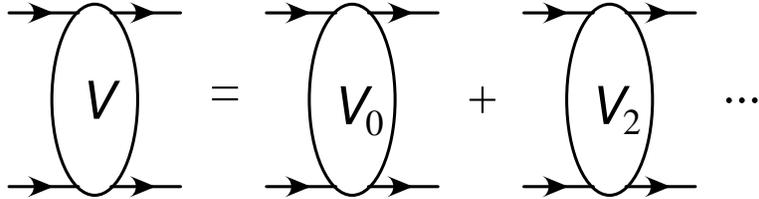}%
\caption{Chiral expansion of the two-particle irreducible kernel.}%
\label{twopi_nn}%
\end{center}
\end{figure}

\begin{figure}
[ptb]
\begin{center}
\includegraphics[
height=1.1225in,
width=5.047in
]%
{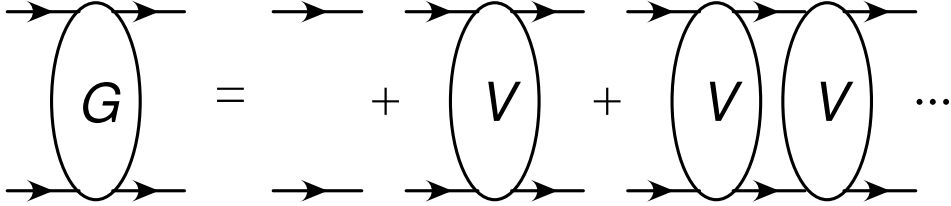}%
\caption{The two-particle irreducible kernel is iterated to all orders to
produce the nucleon-nucleon scattering Green's function.}%
\label{greens_nn}%
\end{center}
\end{figure}

Weinberg showed \cite{Weinberg:1990rz}\cite{Weinberg:1991um} that the simple
diagrammatic expansion for nucleon-nucleon scattering is spoiled by infrared
divergences. He suggested performing an expansion of the two-particle
irreducible kernel (see Fig. \ref{twopi_nn}) and then iterating the kernel to
all orders to produce the scattering Green's function (see Fig.
\ref{greens_nn}). It was later pointed out that a possible difficulty arises
because at any order in $Q$ an infinite number of diagrams is summed, and it
is not clear that all the cutoff dependence at that order can be absorbed into
counterterms that are present at that order \cite{Kaplan:1996xu}. This problem
does indeed arise if one considers nucleon-nucleon scattering in the
$^{1}S_{0}$ channel \cite{Beane:2001bc}, but in practice the cutoff dependence
appears to be very weak [20]\cite{Lepage:1997cs}.%

%

In this work we go one step further and consider the nuclear many-body
problem. We expand the terms in our action order by order,%
\begin{equation}
S=S_{0}+S_{1}+S_{2}+\cdots
\end{equation}
At order $k$ in the chiral expansion, we calculate observables by evaluating
the functional integral%
\begin{equation}
\left\langle G(\bar{N},N,\pi)\right\rangle _{k}=\frac{\int DND\bar{N}D\pi
G(\bar{N},N,\pi)\exp\left[  -S_{0}-S_{1}-\cdots-S_{k}\right]  }{\int
DND\bar{N}D\pi\exp\left[  -S_{0}-S_{1}-\cdots-S_{k}\right]  }.
\end{equation}
We will refer to this approach as non-perturbative effective field theory.
\ The interactions at chiral order $k$ or less are iterated to arbitrary loop
order. \ The functional integral is computed non-perturbatively by putting the
pion and nucleon fields on the lattice and using Monte Carlo sampling. \ Since
the number of diagrams at a given chiral order grows exponentially with the
number of nucleons, a non-perturbative technique such as this is needed for
systems with more than just a few nucleons.

Computing the path integral corresponds to summing an infinite set of
diagrams. As in the case of iterating the two-particle irreducible kernel to
determine the full two-nucleon Green's function, it is not clear that the
cutoff dependence at a given order in the low energy expansion can be absorbed
into a finite number of coefficients in the action. In practice we will
therefore restrict ourselves to lattice cutoffs that satisfy $\pi
a^{-1}<\Lambda_{\chi}$. In order to show that the effective field theory
calculation is consistent we must find a window of lattice cutoffs such that
the many-body calculation is independent of the cutoff up to terms that are
higher order. We shall study this question numerically in our results section.

\section{Lowest order interactions}

Our momentum cutoff scale is $\pi a^{-1}$ and we choose the lattice spacing so
that%
\begin{equation}
m_{\pi}<\pi a^{-1}<\Lambda_{\chi}\text{.}%
\end{equation}
An irreducible diagram is one that cannot be disconnected by cutting internal
lines that match the set of incoming or outgoing particles. \ In the Weinberg
counting scheme \cite{Weinberg:1990rz}\cite{Weinberg:1991um}%
\cite{Weinberg:1992yk} we estimate the chiral order of an irreducible diagram
by associating one power of $Q/\Lambda_{\chi}$ for each derivative interaction
or explicit factor of $m_{\pi}$, four powers for each loop integral, one
inverse power for each nucleon internal line, and two inverse powers for each
pion internal line. \ If $\nu$ is the chiral order of an irreducible diagram,
it can be shown that
\begin{equation}
\nu=4-\tfrac{e_{n}}{2}+2l-2c+\sum_{\text{vertex }i}\delta_{i}. \label{order}%
\end{equation}
In (\ref{order}) $e_{n}$ is the number of external nucleons, $l$ is the number
of loops, $c$ is the number of connected pieces, and $\delta_{i}$ for each
vertex is%
\begin{equation}
\delta_{i}=\#\partial+\#m_{\pi}+\tfrac{\#n}{2}-2,
\end{equation}
where $\#\partial$ is the number of derivatives, $\#m_{\pi}$ is the number of
explicit factors of $m_{\pi}$ in the coefficient, and $\#n$ is the number of
nucleon fields. \ It turns out that $\#m_{\pi}$ is always an even number.

We let $N$ represent the nucleon fields,%
\begin{equation}
N=\left[
\begin{array}
[c]{c}%
\text{proton}\\
\text{neutron}%
\end{array}
\right]  \otimes\left[
\begin{array}
[c]{c}%
\uparrow\\
\downarrow
\end{array}
\right]  .\label{nucfield}%
\end{equation}
We use $\tau_{i}$ to represent Pauli matrices acting in isospin space, and we
use $\vec{\sigma}$ to represent Pauli matrices acting in spin space. Pion
fields are notated as $\pi_{i}$. \ We denote the pion decay constant as
$F_{\pi}^{phys}\approx183$ MeV and let
\begin{equation}
D=1+\pi_{i}^{2}/F_{\pi}^{2}.
\end{equation}
The lowest order Lagrange density for low-energy pions and nucleons is given
by terms with $\delta_{i}=0$ \cite{Ordonez:1996rz},
\begin{align}
\mathcal{L}^{(0)} &  =-\tfrac{1}{2}D^{-2}\left[  (\vec{\nabla}\pi_{i}%
)^{2}-\dot{\pi}_{i}^{2}\right]  -\tfrac{1}{2}D^{-1}m_{\pi}^{2}\pi_{i}^{2}%
+\bar{N}[i\partial_{0}-(m_{N}-\mu)]N\nonumber\\
&  -D^{-1}F_{\pi}^{-1}g_{A}\bar{N}\left[  \tau_{i}\vec{\sigma}\cdot\vec
{\nabla}\pi_{i}\right]  N-D^{-1}F_{\pi}^{-2}\bar{N}[\epsilon_{ijk}\tau_{i}%
\pi_{j}\dot{\pi}_{k}]N\nonumber\\
&  -\tfrac{1}{2}C_{S}:\bar{N}N\bar{N}N:-\tfrac{1}{2}C_{T}:\bar{N}\vec{\sigma
}N\cdot\bar{N}\vec{\sigma}N:.\label{L_0}%
\end{align}
$g_{A}$ is the nucleon axial coupling, and $\epsilon_{ijk}$ is the Levi-Civita
symbol. \ The chemical potential $\mu$ controls the nucleon density and $\mu$
will be set very close to $m_{N}$. \ At next order we have terms with
$\delta_{i}=1$,%
\begin{equation}
\mathcal{L}^{(1)}=\tfrac{1}{2m_{N}}\bar{N}\vec{\nabla}^{2}N+...
\end{equation}
We\ will include this kinetic energy term from $\mathcal{L}^{(1)}$ in our
lowest-order Lagrange density so that we get the usual free nucleon propagator.

In this study we limit ourselves to the interactions of neutrons and neutral
pions and consider processes with only up to two pions. \ As a result we have
at lowest order the terms
\begin{align}
\mathcal{L} &  =-\tfrac{1}{2}\left[  -\dot{\pi}_{0}^{2}+(\vec{\nabla}\pi
_{0})^{2}+m_{\pi}^{2}\pi_{0}^{2}\right]  +a_{j}^{\dagger}[i\partial_{0}%
+\tfrac{\vec{\nabla}^{2}}{2m_{N}}-(m_{N}-\mu)]a_{j}\nonumber\\
&  +\tfrac{g_{A}}{F_{\pi}}a_{i}^{\dagger}\vec{\sigma}_{ij}a_{j}\cdot
\vec{\nabla}\pi_{0}-Ca_{\uparrow}^{\dagger}a_{\uparrow}a_{\downarrow}%
^{\dagger}a_{\downarrow}%
\end{align}
where $a,a^{\dagger}$ are annihilation and creation operators for the neutron.
\ In the Euclidean formalism, we have the partition function%

\begin{equation}
Z=\int D\pi DND\bar{N}\exp\left(  -S_{E}\right)  =\int D\pi DND\bar{N}%
\exp\left(  \int d^{4}x\,\mathcal{L}_{E}\right)
\end{equation}
where%
\begin{align}
\mathcal{L}_{E} &  =-\tfrac{1}{2}\left[  \dot{\pi}_{0}^{2}+(\vec{\nabla}%
\pi_{0})^{2}+m_{\pi}^{2}\pi_{0}^{2}\right]  -a_{j}^{\dagger}[\partial
_{0}-\tfrac{\vec{\nabla}^{2}}{2m_{N}}+(m_{N}-\mu)]a_{j}\nonumber\\
&  +\tfrac{g_{A}}{F_{\pi}}a_{i}^{\dagger}\vec{\sigma}_{ij}a_{j}\cdot
\vec{\nabla}\pi_{0}-Ca_{\uparrow}^{\dagger}a_{\uparrow}a_{\downarrow}%
^{\dagger}a_{\downarrow}.
\end{align}
We will use $x_{0}$ to represent the Euclidean temporal coordinate, rather
than switching from $x_{0}$ to $x_{4}$.

\section{Free neutron}

In the simplest discretization, the Euclidean lattice action for free neutrons
has the form%

\begin{align}
S_{\bar{N}N}^{\text{simple}}  &  =\sum_{\vec{n},i}\left[  c_{i}^{\ast}(\vec
{n})c_{i}(\vec{n}+\hat{0})+\left(  -1+(m_{N}-\mu)\alpha_{t}+6h\right)
c_{i}^{\ast}(\vec{n})c_{i}(\vec{n})\right] \nonumber\\
&  -h\sum_{\vec{n},\hat{l}_{s},i}\left[  c_{i}^{\ast}(\vec{n})c_{i}(\vec
{n}+\hat{l}_{s})+c_{i}^{\ast}(\vec{n})c_{i}(\vec{n}-\hat{l}_{s})\right]
\label{simplefree}%
\end{align}
where%
\begin{equation}
h=\tfrac{\alpha_{t}}{2m_{N}}.
\end{equation}
However we want a discretization that will minimize the dependence on
$\alpha_{t}$ so that fewer lattice steps in the temporal direction can be used
and results for different $\alpha_{t}$ can be directly compared.

Let us review the conversion from the operator formalism to path integrals.
\ The free neutron lattice Hamiltonian is%
\begin{align}
H_{\bar{N}N}  &  =\sum_{\vec{n}_{s},i}\left[  (m_{N}-\mu+\tfrac{3}{m_{N}%
})a_{i}^{\dagger}(\vec{n}_{s})a_{i}(\vec{n}_{s})\right] \nonumber\\
&  -\tfrac{1}{2m_{N}}\sum_{\vec{n}_{s},\hat{l}_{s},i}\left[  a_{i}^{\dagger
}(\vec{n}_{s})a_{i}(\vec{n}_{s}+\hat{l}_{s})+a_{i}^{\dagger}(\vec{n}_{s}%
)a_{i}(\vec{n}_{s}-\hat{l}_{s})\right]  .
\end{align}
We want to convert the partition function for free neutrons,%
\begin{equation}
Z=\text{Tr}\left[  \exp(-\beta H_{\bar{N}N})\right]  =\text{Tr}\left[
\exp(-\alpha_{t}H_{\bar{N}N})\exp(-\alpha_{t}H_{\bar{N}N})...\exp(-\alpha
_{t}H_{\bar{N}N})\right]  ,
\end{equation}
into the form%

\begin{equation}
Z=\int DcDc^{\ast}\exp\left[  -S_{\bar{N}N}\right]  . \label{basicpath}%
\end{equation}
Using the identity \cite{Creutz:1999zy}%
\begin{equation}
\exp[a_{i}^{\dagger}X_{ij}a_{j}]=:\exp[a_{i}^{\dagger}(e^{X}-1)_{ij}a_{j}]:,
\end{equation}
we can write%
\begin{equation}
\exp(-\alpha_{t}H_{\bar{N}N})=:\exp(-h_{\bar{N}N}(a^{\dagger},a)):+O(h),
\end{equation}
where%
\begin{align}
h_{\bar{N}N}(a^{\dagger},a)  &  =\sum_{\vec{n}_{s},i}\left[  (1-e^{-((m_{N}%
-\mu)\alpha_{t}+6h)})a_{i}^{\dagger}(\vec{n}_{s})a_{i}(\vec{n}_{s})\right]
\nonumber\label{op}\\
&  -he^{-((m_{N}-\mu)\alpha_{t})}\sum_{\vec{n}_{s},\hat{l}_{s},i}\left[
a_{i}^{\dagger}(\vec{n}_{s})a_{i}(\vec{n}_{s}+\hat{l}_{s})+a_{i}^{\dagger
}(\vec{n}_{s})a_{i}(\vec{n}_{s}-\hat{l}_{s})\right]  .
\end{align}
Introducing the extra $e^{-((m_{N}-\mu)\alpha_{t})}$ factor multiplying $h$ is
not well motivated at this stage, but it insures that the neutron chemical
potential is coupled to an exactly conserved neutron number operator
\cite{Hasenfratz:1983ba}\cite{Hasenfratz:1984em}. \ We now use the
correspondence \cite{Creutz:1999zy}\cite{Creutz:1988wv}%
\begin{align}
&  \text{Tr}\left[  :f_{n-1}(a^{\dagger},a):...:f_{1}(a^{\dagger}%
,a)::f_{0}(a^{\dagger},a):\right] \nonumber\\
&  =\int dc_{n-1}dc_{n-1}^{\ast}...dc_{0}dc_{0}^{\ast}\exp[\sum_{j=0,...,n-1}%
c_{j}^{\ast}(c_{j}-c_{j+1})]\prod_{j=0,...,n-1}f_{j}(c_{j}^{\ast},c_{j})
\end{align}
with $c_{n}=-c_{0}$. \ We can now convert the partition function to the path
integral form in (\ref{basicpath}) with%
\begin{align}
S_{\bar{N}N}  &  =\sum_{\vec{n}_{s},i}\left[  c_{i}^{\ast}(\vec{n})c_{i}%
(\vec{n}+\hat{0})-e^{-((m_{N}-\mu)\alpha_{t}+6h)}c_{i}^{\ast}(\vec{n}%
)c_{i}(\vec{n})\right] \nonumber\\
&  -he^{-(m_{N}-\mu)\alpha_{t}}\sum_{\vec{n},l_{s}i}\left[  c_{i}^{\ast}%
(\vec{n})c_{i}(\vec{n}+\hat{l}_{s})+c_{i}^{\ast}(\vec{n})c_{i}(\vec{n}-\hat
{l}_{s})\right]  .
\end{align}
This lattice action has temporal discretization errors of $O(h)$, whereas the
action in (\ref{simplefree}) has errors of $O(\alpha_{t})$. \ Since $h$ is a
small parameter this is an improvement and so the dependence on $\alpha_{t}$
has been significantly reduced.

It is conventional to define a new normalization for $c_{i}$,
\begin{equation}
c_{i}^{\prime}=c_{i}e^{-(m_{N}-\mu)\alpha_{t}}.
\end{equation}
Then
\begin{equation}
Z=e^{-2(m_{N}-\mu)\beta L^{3}}\int Dc^{\prime}Dc^{\ast}\exp\left[  -S_{\bar
{N}N}\right]
\end{equation}
where%
\begin{align}
S_{\bar{N}N}  &  =\sum_{\vec{n},i}\left[  e^{(m_{N}-\mu)\alpha_{t}}c_{i}%
^{\ast}(\vec{n})c_{i}^{\prime}(\vec{n}+\hat{0})-e^{-6h}c_{i}^{\ast}(\vec
{n})c_{i}^{\prime}(\vec{n})\right] \nonumber\\
&  -h\sum_{\vec{n},\hat{l}_{s},i}\left[  c_{i}^{\ast}(\vec{n})c_{i}^{\prime
}(\vec{n}+\hat{l}_{s})+c_{i}^{\ast}(\vec{n})c_{i}^{\prime}(\vec{n}-\hat{l}%
_{s})\right]  . \label{improvedfree}%
\end{align}
We observe that the neutron chemical potential is in fact coupled to an
exactly conserved neutron number operator since it appears in the same manner
as a lattice gauge connection in the temporal direction. \ Comparing the two
actions (\ref{simplefree}) and (\ref{improvedfree}) we can summarize the
difference as follows. \ If we write%

\begin{align}
S_{\bar{N}N}^{\text{simple}}  &  =\sum_{\vec{n},i}\left[  c_{i}^{\ast}(\vec
{n})c_{i}(\vec{n}+\hat{0})\right]  +\sum_{\vec{n},i}\left[  \left(
-1+(m_{N}-\mu)\alpha_{t}+X\right)  c_{i}^{\ast}(\vec{n})c_{i}(\vec{n})\right]
\nonumber\\
&  -h\sum_{\vec{n},\hat{l}_{s},i}\left[  c_{i}^{\ast}(\vec{n})c_{i}(\vec
{n}+\hat{l}_{s})+c_{i}^{\ast}(\vec{n})c_{i}(\vec{n}-\hat{l}_{s})\right]
\end{align}
where%

\begin{equation}
X=6h,
\end{equation}
then%
\begin{align}
S_{\bar{N}N}  &  =\sum_{\vec{n},i}\left[  e^{(m_{N}-\mu)\alpha_{t}}c_{i}%
^{\ast}(\vec{n})c_{i}^{\prime}(\vec{n}+\hat{0})\right]  -\sum_{\vec{n}%
,i}\left[  c_{i}^{\ast}(\vec{n})e^{-X}c_{i}^{\prime}(\vec{n})\right]
\nonumber\\
&  -h\sum_{\vec{n},\hat{l}_{s},i}\left[  c_{i}^{\ast}(\vec{n})c_{i}^{\prime
}(\vec{n}+\hat{l}_{s})+c_{i}^{\ast}(\vec{n})c_{i}^{\prime}(\vec{n}-\hat{l}%
_{s})\right]  .
\end{align}

In momentum space we have%
\begin{equation}
S_{\bar{N}N}=\sum_{\vec{k},i}\tilde{c}_{i}^{\ast}(-\vec{k})\tilde{c}%
_{i}^{\prime}(\vec{k})\left[  e^{-ik_{\ast0}+(m_{N}-\mu)\alpha_{t}}%
-e^{-6h}-2h\sum_{\hat{l}_{s}}\cos(k_{\ast l_{s}})\right]  .
\end{equation}
We now have the free neutron correlation function on the lattice,%
\begin{equation}
\frac{\int Dc^{\prime}Dc^{\ast}c_{i}^{\prime}(\vec{n})c_{i}^{\ast}%
(0)\exp\left[  -S_{\bar{N}N}\right]  }{\int Dc^{\prime}Dc^{\ast}\exp\left[
-S_{\bar{N}N}\right]  }=\frac{1}{L_{t}L^{3}}\sum_{\vec{k}}e^{-i\vec{k}_{\ast
}\cdot\vec{n}}D_{N}(\vec{k}), \label{pathcorrelator}%
\end{equation}
(no sum over $i$) where the free neutron propagator is%
\begin{equation}
D_{N}(\vec{k})=\frac{1}{e^{-ik_{\ast0}+(m_{N}-\mu)\alpha_{t}}-e^{-6h}%
-2h\sum_{\hat{l}_{s}}\cos(k_{\ast l_{s}})}.
\end{equation}

\section{Neutron correlation functions}

At nonzero time step there are some subtleties going from the correlation
functions in the operator formalism to correlation functions in the path
integral formalism. \ \ We have%

\begin{align}
Z  &  =\text{Tr}\left[  \exp(-\alpha_{t}H)...\exp(-\alpha_{t}H)\exp
(-\alpha_{t}H)\right] \nonumber\\
&  =\text{Tr}\left[  :f_{L_{t}-1}(a^{\dagger},a):\cdots:f_{1}(a^{\dagger
},a)::f_{0}(a^{\dagger},a):\right]
\end{align}
where the total number of time steps is $L_{t}$ and each of the $f_{j}$'s are
the same,%
\begin{equation}
:f_{j}(a^{\dagger},a):=\exp(-\alpha_{t}H)+O(h).
\end{equation}
Suppose we wish to calculate%
\begin{equation}
\text{Tr}\left[  u_{L_{t}-1}(a)v_{L_{t}-1}(a^{\dagger}):f_{L_{t}-1}:\cdots
u_{1}(a)v_{1}(a^{\dagger}):f_{1}:u_{0}(a)v_{0}(a^{\dagger}):f_{0}:\right]  .
\end{equation}
This can be rewritten as%

\begin{equation}
\text{Tr}\left[  :v_{L_{t}-1}(a^{\dagger})f_{L_{t}-1}u_{L_{t}-2}%
(a):\cdots:v_{1}(a^{\dagger})f_{1}u_{0}(a)::v_{0}(a^{\dagger})f_{0}u_{L_{t}%
-1}(a):\right]  .
\end{equation}
In the path integral formalism this is equivalent to%
\begin{equation}
\int DcDc^{\ast}F(c,c^{\ast})\exp[-S]
\end{equation}
where%
\begin{equation}
F(c,c^{\ast})=v_{L_{t}-1}(c_{L_{t}-1}^{\ast})u_{L_{t}-2}(c_{L_{t}-1})\cdots
v_{0}(c_{0}^{\ast})u_{L_{t}-1}(c_{0}).
\end{equation}
We note that $v_{j}$ is a function of $c_{j}^{\ast}$ whereas $u_{j}$ is a
function of $c_{j+1}^{\ast}$.

\section{Free neutral pion}

The lattice action for a free neutral pion is%
\begin{equation}
S_{\pi\pi}=((\tfrac{m_{\pi}^{2}}{2}+3)\alpha_{t}+\alpha_{t}^{-1})\sum_{\vec
{n}}\pi(\vec{n})\pi(\vec{n})-\sum_{\vec{n},\hat{l}}\left[  e_{l}\pi(\vec
{n})\pi(\vec{n}+\hat{l})\right]
\end{equation}
where%
\begin{equation}
(e_{0},e_{1},e_{2},e_{3})=(\alpha_{t}^{-1},\alpha_{t},\alpha_{t},\alpha_{t}).
\end{equation}
In momentum space the action is%
\begin{equation}
S_{\pi\pi}=\sum_{\vec{n}}\pi(-\vec{k})\pi(\vec{k})\left[  (\tfrac{m_{\pi}^{2}%
}{2}+3)\alpha_{t}+\alpha_{t}^{-1}-\sum_{\hat{l}}e_{l}\cos(k_{\ast l})\right]
\end{equation}
and so
\begin{equation}
\frac{\int D\pi\pi(\vec{n})\pi(0)\exp\left[  -S_{\pi\pi}\right]  }{\int
D\pi\exp\left[  -S_{\pi\pi}\right]  }=\frac{1}{L_{t}L^{3}}\sum_{\vec{k}%
}e^{-i\vec{k}_{\ast}\cdot\vec{n}}D_{\pi}(\vec{k}),
\end{equation}
where free neutral pion propagator is%
\begin{equation}
D_{\pi}(\vec{k})=\frac{1}{2\left[  (\tfrac{m_{\pi}^{2}}{2}+3)\alpha_{t}%
+\alpha_{t}^{-1}-\sum_{\hat{l}}e_{l}\cos(k_{\ast l})\right]  }.
\end{equation}

In this first exploratory study we are not concerned with the issue of exact
chiral symmetry on the lattice and therefore will neglect the Haar measure.
\ This aspect of exact chiral symmetry will be investigated in a future study
along with the inclusion of charged nucleons and pions.

\section{Neutral pion-neutron coupling}

In the continuum the pion-nucleon coupling makes a contribution to the
integrand of the partition function of the form%
\begin{equation}
\exp\left[  -\tfrac{g_{A}}{F_{\pi}}\int d^{4}x\;D^{-1}\bar{N}[\tau_{a}%
(\vec{\sigma}\cdot\vec{\nabla}\pi_{a})]N\right]  ,
\end{equation}
where $\vec{\sigma}=(\sigma_{1},\sigma_{2},\sigma_{3})$ are Pauli matrices for
spin, $\tau_{a}=(\tau_{1},\tau_{2},\tau_{3})$ are Pauli matrices for isospin,%
\begin{equation}
D=1+\tfrac{\vec{\pi}^{2}}{F_{\pi}^{2}}\text{,}%
\end{equation}
and $F_{\pi}^{phys}\approx183$ MeV is the pion decay constant,%
\begin{equation}
\left\langle 0\right\vert j_{5a}^{\mu}\left\vert \pi_{b}(p)\right\rangle
=i\delta_{ab}\tfrac{F_{\pi}}{2}p^{\mu}.
\end{equation}

We keep only the term involving the neutral pion and neutron,%
\begin{equation}
\exp\left[  \tfrac{g_{A}}{F_{\pi}}\int d^{4}x\;\left(  c_{i}^{\ast}\vec
{\sigma}_{ij}c_{j}\cdot\vec{\nabla}\pi_{0}\right)  \right]  .
\end{equation}
The simplest lattice discretization of this interaction term is%

\begin{equation}
\exp\left[  -S_{\pi\bar{N}N}\right]
\end{equation}
where%

\begin{align}
S_{\pi\bar{N}N}^{\text{simple}}  &  =-\tfrac{g_{A}\alpha_{t}}{2F_{\pi}}%
{\displaystyle\sum_{\vec{n}}}
\left[  \left[  c_{\uparrow}^{\ast}(\vec{n})c_{\uparrow}(\vec{n}%
)-c_{\downarrow}^{\ast}(\vec{n})c_{\downarrow}(\vec{n})\right]  \Delta
_{3}^{\pm}\pi_{0}(\vec{n})\right] \nonumber\\
&  -\tfrac{g_{A}\alpha_{t}}{2F_{\pi}}%
{\displaystyle\sum_{\vec{n}}}
\left[  c_{\uparrow}^{\ast}(\vec{n})c_{\downarrow}(\vec{n})\left[  \Delta
_{1}^{\pm}\pi_{0}(\vec{n})-i\Delta_{2}^{\pm}\pi_{0}(\vec{n})\right]  \right]
\nonumber\\
&  -\tfrac{g_{A}\alpha_{t}}{2F_{\pi}}%
{\displaystyle\sum_{\vec{n}}}
\left[  c_{\downarrow}^{\ast}(\vec{n})c_{\uparrow}(\vec{n})\left[  \Delta
_{1}^{\pm}\pi_{0}(\vec{n})+i\Delta_{2}^{\pm}\pi_{0}(\vec{n})\right]  \right]
.
\end{align}
and
\begin{equation}
\Delta_{l}^{\pm}\pi_{0}(\vec{n})=\pi_{0}(\vec{n}+\hat{l})-\pi_{0}(\vec{n}%
-\hat{l}).
\end{equation}

We now use a temporally-improved discretization. \ We can write the simple
lattice action for the free neutron with pion-neutron coupling as
\begin{align}
S_{\bar{N}N}^{\text{simple}}+S_{\pi\bar{N}N}^{\text{simple}}  &  =\sum
_{\vec{n},i}c_{i}^{\ast}(\vec{n})c_{i}(\vec{n}+\hat{0})\nonumber\\
&  +\sum_{\vec{n},i,j}c_{i}^{\ast}(\vec{n})\left(  [-1+(m_{N}-\mu)\alpha
_{t}]\delta_{ij}+X_{ij}(\vec{n})\right)  c_{j}(\vec{n})\nonumber\\
&  -h\sum_{\vec{n},\hat{l}_{s},i}\left[  c_{i}^{\ast}(\vec{n})c_{i}(\vec
{n}+\hat{l}_{s})+c_{i}^{\ast}(\vec{n})c_{i}(\vec{n}-\hat{l}_{s})\right]
\end{align}
with $X_{ij}(\vec{n})$ given by the matrix%
\begin{equation}
\left[
\begin{array}
[c]{cc}%
6h-\tfrac{g_{A}\alpha_{t}}{2F_{\pi}}\Delta_{3}^{\pm}\pi_{0}(\vec{n}) &
-\tfrac{g_{A}\alpha_{t}}{2F_{\pi}}\left(  \Delta_{1}^{\pm}-i\Delta_{2}^{\pm
}\right)  \pi_{0}(\vec{n})\\
-\tfrac{g_{A}\alpha_{t}}{2F_{\pi}}\left(  \Delta_{1}^{\pm}+i\Delta_{2}^{\pm
}\right)  \pi_{0}(\vec{n}) & 6h+\tfrac{g_{A}\alpha_{t}}{2F_{\pi}}\Delta
_{3}^{\pm}\pi_{0}(\vec{n})
\end{array}
\right]
\end{equation}
Then our temporally-improved action is%
\begin{align}
S_{\bar{N}N+\pi\bar{N}N}  &  =\sum_{\vec{n},i}e^{(m_{N}-\mu)\alpha_{t}}%
c_{i}^{\ast}(\vec{n})c_{i}^{\prime}(\vec{n}+\hat{0})\nonumber\\
&  -\sum_{\vec{n},i,j}c_{i}^{\ast}(\vec{n})\left(  e^{-X(\vec{n})}\right)
_{ij}c_{j}^{\prime}(\vec{n})\nonumber\\
&  -h\sum_{\vec{n},\hat{l}_{s},i}\left[  c_{i}^{\ast}(\vec{n})c_{i}^{\prime
}(\vec{n}+\hat{l}_{s})+c_{i}^{\ast}(\vec{n})c_{i}^{\prime}(\vec{n}-\hat{l}%
_{s})\right]  .
\end{align}

\section{Neutron contact term}

The neutron contact term has the form%
\begin{equation}
H_{\bar{N}N\bar{N}N}=C\sum_{\vec{n},i}a_{\uparrow}^{\dagger}(\vec
{n})a_{\uparrow}(\vec{n})a_{\downarrow}^{\dagger}(\vec{n})a_{\downarrow}%
(\vec{n}).
\end{equation}
We can rewrite the contribution at lattice site $\vec{n}$ to the partition
function using a discrete Hubbard-Stratonovich transformation
\cite{Hirsch:1983}. \ For $C\leq0,$%
\begin{align}
&  \exp(-C\alpha_{t}a_{\uparrow}^{\dagger}(\vec{n})a_{\uparrow}(\vec
{n})a_{\downarrow}^{\dagger}(\vec{n})a_{\downarrow}(\vec{n}))\nonumber\\
&  =\tfrac{1}{2}\sum_{s(\vec{n})=\pm1}\exp\left[  -(\tfrac{C\alpha_{t}}%
{2}+\lambda s(\vec{n}))(a_{\uparrow}^{\dagger}(\vec{n})a_{\uparrow}(\vec
{n})+a_{\downarrow}^{\dagger}(\vec{n})a_{\downarrow}(\vec{n})-1)\right]  ,
\label{operator}%
\end{align}
where%
\begin{equation}
\cosh\lambda=\exp(-\tfrac{C\alpha_{t}}{2}).
\end{equation}
Since%
\begin{equation}
e^{\lambda}+e^{-\lambda}=2\exp(-\tfrac{C\alpha_{t}}{2}),
\end{equation}
we can write%
\begin{equation}
\lambda=\ln\left[  \exp(-\tfrac{C\alpha_{t}}{2})+\sqrt{\exp(-C\alpha_{t}%
)-1}\right]  .
\end{equation}
The simplest lattice discretization gives a contribution to action,%

\begin{equation}
-\tfrac{C\beta}{2}L^{3}+S_{ss}+S_{s\bar{N}N}^{\text{simple}}%
\end{equation}
where%
\begin{equation}
S_{ss}=-\sum_{\vec{n}}\lambda s(\vec{n})
\end{equation}
and%
\begin{equation}
S_{s\bar{N}N}^{\text{simple}}=\sum_{\vec{n}}\left[  (\tfrac{C\alpha_{t}}%
{2}+\lambda s(\vec{n}))(c_{\uparrow}^{\ast}(\vec{n})c_{\uparrow}(\vec
{n})+c_{\downarrow}^{\ast}(\vec{n})c_{\downarrow}(\vec{n}))\right]  .
\end{equation}
However this actually gives a result that is inconsistent with the Hamiltonian
operator form in (\ref{operator}) in limit $\alpha_{t}\rightarrow0$. \ The
problem is somewhat subtle and has not been given much attention in the
literature. \ The point is that operator ordering at $O(\lambda^{2})$ cannot
be ignored since $\lambda^{2}\sim O(\alpha_{t})$. \ We will deal with this in
the same way that we constructed the temporally-improved action for the free
neutron and the pion-neutron coupling. \ We write%
\begin{align}
&  S_{\bar{N}N}^{\text{simple}}+S_{\pi\bar{N}N}^{\text{simple}}+S_{s\bar{N}%
N}^{\text{simple}}\nonumber\\
&  =\sum_{\vec{n},i}c_{i}^{\ast}(\vec{n})c_{i}(\vec{n}+\hat{0})\nonumber\\
&  +\sum_{\vec{n},i,j}c_{i}^{\ast}(\vec{n})\left(  [-1+(m_{N}-\mu)\alpha
_{t}]\delta_{ij}+X_{ij}(\vec{n})\right)  c_{j}(\vec{n})\nonumber\\
&  -h\sum_{\vec{n},\hat{l}_{s},i}\left[  c_{i}^{\ast}(\vec{n})c_{i}(\vec
{n}+\hat{l}_{s})+c_{i}^{\ast}(\vec{n})c_{i}(\vec{n}-\hat{l}_{s})\right]
\end{align}
with~$X_{ij}(\vec{n})$ equal to%
\begin{equation}
\left[
\begin{array}
[c]{cc}%
6h-\tfrac{g_{A}\alpha_{t}}{2F_{\pi}}\Delta_{3}^{\pm}\pi_{0}(\vec{n}%
)+\tfrac{C\alpha_{t}}{2}+\lambda s(\vec{n}) & -\tfrac{g_{A}\alpha_{t}}%
{2F_{\pi}}\left(  \Delta_{1}^{\pm}-i\Delta_{2}^{\pm}\right)  \pi_{0}(\vec
{n})\\
-\tfrac{g_{A}\alpha_{t}}{2F_{\pi}}\left(  \Delta_{1}^{\pm}+i\Delta_{2}^{\pm
}\right)  \pi_{0}(\vec{n}) & 6h+\tfrac{g_{A}\alpha_{t}}{2F_{\pi}}\Delta
_{3}^{\pm}\pi_{0}(\vec{n})+\tfrac{C\alpha_{t}}{2}+\lambda s(\vec{n})
\end{array}
\right]
\end{equation}
then%
\begin{align}
S_{\bar{N}N+\pi\bar{N}N+s\bar{N}N}  &  =\sum_{\vec{n},i}e^{(m_{N}-\mu
)\alpha_{t}}c_{i}^{\ast}(\vec{n})c_{i}^{\prime}(\vec{n}+\hat{0})\nonumber\\
&  -\sum_{\vec{n},i,j}c_{i}^{\ast}(\vec{n})\left(  e^{-X(\vec{n})}\right)
_{ij}c_{j}^{\prime}(\vec{n})\nonumber\\
&  -h\sum_{\vec{n},\hat{l}_{s},i}\left[  c_{i}^{\ast}(\vec{n})c_{i}(\vec
{n}+\hat{l}_{s})+c_{i}^{\ast}(\vec{n})c_{i}(\vec{n}-\hat{l}_{s})\right]  .
\end{align}

\section{One-loop neutron self-energy}

In the next few sections we calculate several lattice Feynman diagrams.
\ These calculations will serve as a check that our non-perturbative
simulation is functioning properly in the small coupling limit $g_{A}%
\rightarrow0$, $C\rightarrow0$. It will also give us a reference point to
measure how non-perturbative the interactions are at physical values for
$g_{A}$ and $C$ and for various densities.%

\begin{figure}
[ptb]
\begin{center}
\includegraphics[
height=1.516in,
width=3.039in
]%
{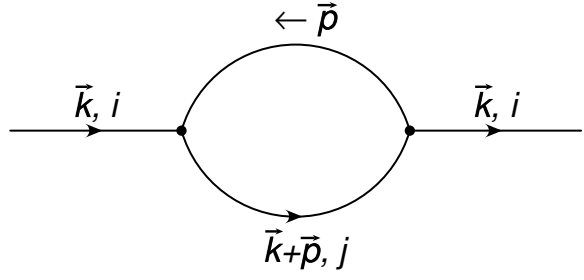}%
\caption{Neutron self-energy due to a neutral pion and neutron intermediate
states.}%
\label{n_self_n_pi}%
\end{center}
\end{figure}

At $O(g_{A}^{2})$ we have a contribution to the neutron self-energy due to a
neutral pion and neutron intermediate state as shown in Fig. \ref{n_self_n_pi}%
. If we write the pion-interaction interaction term to order $O(g_{A})$ in
momentum space, we have a contribution to the action of the form%
\begin{align}
&  \tfrac{ig_{A}\alpha_{t}e^{-6h}}{F_{\pi}\sqrt{L^{3}L_{t}}}\sum_{\vec{k}%
,\vec{p}}\left[  \tilde{c}_{\uparrow}^{\ast}(-\vec{k}-\vec{p})\tilde
{c}_{\uparrow}^{\prime}(\vec{k})-\tilde{c}_{\downarrow}^{\ast}(-\vec{k}%
-\vec{p})\tilde{c}_{\downarrow}^{\prime}(\vec{k})\right]  \pi_{0}(\vec{p}%
)\sin(p_{\ast3})\nonumber\\
&  +\tfrac{ig_{A}\alpha_{t}e^{-6h}}{F_{\pi}\sqrt{L^{3}L_{t}}}\sum_{\vec
{k},\vec{p}}\left[  \tilde{c}_{\uparrow}^{\ast}(-\vec{k}-\vec{p})\tilde
{c}_{\downarrow}^{\prime}(\vec{k})\right]  \pi_{0}(\vec{p})\left[
\sin(p_{\ast1})-i\sin(p_{\ast2})\right] \nonumber\\
&  +\tfrac{ig_{A}\alpha_{t}e^{-6h}}{F_{\pi}\sqrt{L^{3}L_{t}}}\sum_{\vec
{k},\vec{p}}\left[  \tilde{c}_{\downarrow}^{\ast}(-\vec{k}-\vec{p})\tilde
{c}_{\uparrow}^{\prime}(\vec{k})\right]  \pi_{0}(\vec{p})\left[  \sin
(p_{\ast1})+i\sin(p_{\ast2})\right]  .
\end{align}
Then the diagram in Fig. \ref{n_self_n_pi} leads to a contribution to the
self-energy that goes as%
\begin{equation}
\Sigma_{N}^{(g_{A}^{2})}(\vec{k})=\tfrac{g_{A}^{2}\alpha_{t}^{2}e^{-12h}%
}{F_{\pi}^{2}L^{3}L_{t}}\sum_{\vec{p}}D_{\pi}(\vec{p})D_{N}(\vec{k}+\vec
{p})\left[  \sin^{2}(p_{\ast1})+\sin^{2}(p_{\ast2})+\sin^{2}(p_{\ast
3})\right]  . \label{neutronselfg2}%
\end{equation}
%

\begin{figure}
[ptb]
\begin{center}
\includegraphics[
height=1.1416in,
width=3.039in
]%
{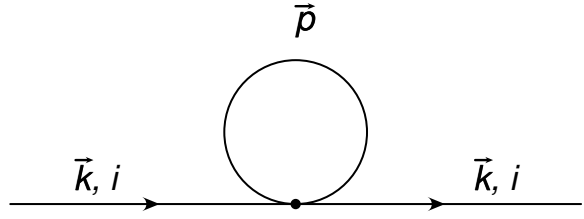}%
\caption{Neutron self-energy due to the $\pi\pi\bar{N}N$ interaction in the
temporally-improved action.}%
\label{n_self_improved}%
\end{center}
\end{figure}

Our temporally-improved action has a $\pi\pi\bar{N}N$ interaction of the form%

\begin{equation}
S=...-\tfrac{g_{A}^{2}\alpha_{t}^{2}}{8F_{\pi}^{2}}e^{-6h}\sum_{\vec{n}%
,\hat{l}_{s},i}c_{i}^{\ast}(\vec{n})c_{i}^{\prime}(\vec{n})(\Delta_{l_{s}%
}^{\pm}\pi_{0}(n))^{2},
\end{equation}
which gives rise to the diagram in Fig. \ref{n_self_improved}. We get an
additional contribution to the self-energy,%

\begin{equation}
\Sigma_{N}^{(g_{A}^{2})}(\vec{k})=\cdots+\tfrac{g_{A}^{2}\alpha_{t}^{2}%
}{8F_{\pi}^{2}}e^{-6h}X, \label{improveneutron}%
\end{equation}
where%
\begin{equation}
X=\frac{4}{L_{t}L^{3}}\sum_{\vec{p}}D_{\pi}(\vec{p})\left[  \sin^{2}(p_{\ast
1})+\sin^{2}(p_{\ast2})+\sin^{2}(p_{\ast3})\right]  . \label{defx}%
\end{equation}

\begin{figure}
[ptb]
\begin{center}
\includegraphics[
height=1.1416in,
width=3.039in
]%
{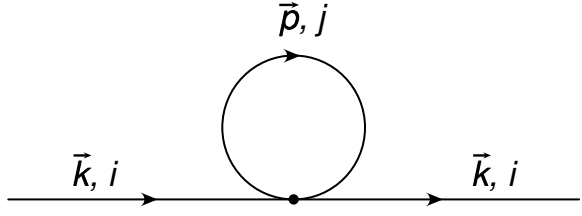}%
\caption{Neutron self-energy due to the contact interaction.}%
\label{n_self_contact}%
\end{center}
\end{figure}

At $O(C)$ we have the one-loop diagram shown in Fig. \ref{n_self_contact}.
If the vertex is located at lattice site $\vec{n}$ we can isolate the relevant
lowest-order interaction in the path integral starting from%
\begin{equation}
\tfrac{1}{2}\sum_{s(\vec{n})=\pm1}e^{\left(  C\alpha_{t}/2+\lambda s(\vec
{n})\right)  }\exp\left[  e^{-6h}\left(  e^{-\left(  C\alpha_{t}/2+\lambda
s(\vec{n})\right)  }-1\right)  \left(
\begin{array}
[c]{c}%
c_{\uparrow}^{\ast}(\vec{n})c_{\uparrow}^{\prime}(\vec{n})\\
+c_{\downarrow}^{\ast}(\vec{n})c_{\downarrow}^{\prime}(\vec{n})
\end{array}
\right)  \right]  .
\end{equation}
Expanding the exponential we get%
\begin{equation}
\tfrac{1}{2}\sum_{s(\vec{n})=\pm1}e^{\left(  C\alpha_{t}/2+\lambda s(\vec
{n})\right)  }\left\{
\begin{array}
[c]{c}%
1+e^{-6h}\left(  e^{-\left(  C\alpha_{t}/2+\lambda s(\vec{n})\right)
}-1\right)  \left(
\begin{array}
[c]{c}%
c_{\uparrow}^{\ast}(\vec{n})c_{\uparrow}^{\prime}(\vec{n})\\
+c_{\downarrow}^{\ast}(\vec{n})c_{\downarrow}^{\prime}(\vec{n})
\end{array}
\right) \\
+e^{-12h}\left(  e^{-\left(  C\alpha_{t}/2+\lambda s(\vec{n})\right)
}-1\right)  ^{2}c_{\uparrow}^{\ast}(\vec{n})c_{\uparrow}^{\prime}(\vec
{n})c_{\downarrow}^{\ast}(\vec{n})c_{\downarrow}^{\prime}(\vec{n})
\end{array}
\right\}  .
\end{equation}
We find that%

\begin{equation}
\tfrac{1}{2}\sum_{s(\vec{n})=\pm1}e^{\left(  C\alpha_{t}/2+\lambda s(\vec
{n})\right)  }\left(  e^{-\left(  C\alpha_{t}/2+\lambda s(\vec{n})\right)
}-1\right)  =0 \label{quadvanish}%
\end{equation}
and%
\begin{equation}
\tfrac{1}{2}\sum_{s(\vec{n})=\pm1}e^{\left(  C\alpha_{t}/2+\lambda s(\vec
{n})\right)  }\left(  e^{-\left(  C\alpha_{t}/2+\lambda s(\vec{n})\right)
}-1\right)  ^{2}=e^{-C\alpha_{t}}-1. \label{quartic}%
\end{equation}
So to lowest order we can write the interaction as%
\begin{equation}
e^{-12h}\left(  e^{-C\alpha_{t}}-1\right)  c_{\uparrow}^{\ast}(\vec
{n})c_{\uparrow}^{\prime}(\vec{n})c_{\downarrow}^{\ast}(\vec{n})c_{\downarrow
}^{\prime}(\vec{n}). \label{effective}%
\end{equation}
Therefore the contribution to the self-energy is%
\begin{equation}
\Sigma_{N}^{(C)}(\vec{k})=e^{-12h}(e^{-C\alpha_{t}}-1)Y, \label{self_c}%
\end{equation}
where%
\begin{align}
Y  &  =\frac{\int Dc^{\prime}Dc^{\ast}c_{i}^{\ast}(0)c_{i}^{\prime}%
(0)\exp\left[  -S_{\bar{N}N}\right]  }{\int Dc^{\prime}Dc^{\ast}\exp\left[
-S_{\bar{N}N}\right]  }\text{ (no sum over }i\text{)}\nonumber\\
&  =-\frac{1}{L_{t}L^{3}}\sum_{\vec{k}}D_{N}(\vec{k}). \label{defy}%
\end{align}
This self-energy term is proportional to $Y$, which in turn is proportional to
the neutron density with an $O(\alpha_{t})$ time discretization correction.

\section{One-loop pion self-energy}%

\begin{figure}
[ptb]
\begin{center}
\includegraphics[
height=1.516in,
width=3.039in
]%
{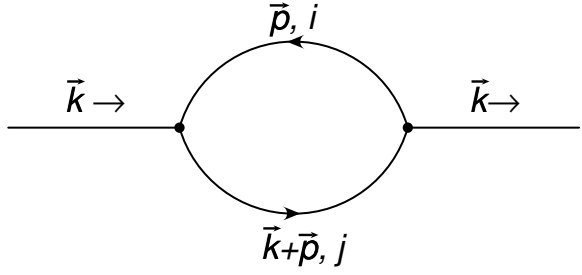}%
\caption{Pion self-energy due to neutron-neutron hole intermediate states.}%
\label{pi_self}%
\end{center}
\end{figure}

\begin{figure}
[ptb]
\begin{center}
\includegraphics[
height=1.1416in,
width=3.039in
]%
{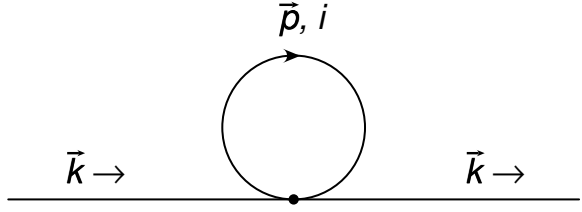}%
\caption{Pion self-energy due to the $\pi\pi N\bar{N}$ interaction in the
temporally-improved action.}%
\label{pi_self_improved}%
\end{center}
\end{figure}

At $O(g_{A}^{2})$ we have a contribution to the pion self-energy due to a
neutron and neutron-hole intermediate state as shown in Fig. \ref{pi_self}.
The contribution to the self-energy is%

\begin{equation}
\Sigma_{\pi}^{(g_{A}^{2})}(\vec{k})=-\tfrac{2g_{A}^{2}\alpha_{t}^{2}e^{-12h}%
}{F_{\pi}^{2}L^{3}L_{t}}\sum_{\vec{p}}D_{N}(\vec{p})D_{N}(\vec{k}+\vec
{p})\left[  \sin^{2}(k_{\ast1})+\sin^{2}(k_{\ast2})+\sin^{2}(k_{\ast
3})\right]  , \label{piselfg2}%
\end{equation}
Our temporally-improved action has a term of the form%
\begin{equation}
S=\cdots-\tfrac{g_{A}^{2}\alpha_{t}^{2}}{8F_{\pi}^{2}}e^{-6h}\sum_{\vec
{n},\hat{l}_{s},i}c_{i}^{\ast}(\vec{n})c_{i}^{\prime}(\vec{n})(\Delta_{l_{s}%
}^{\pm}\pi_{0}(n))^{2}.
\end{equation}
This leads to the diagram in Fig. \ref{pi_self_improved}%
and gives an additional contribution
\begin{equation}
\Sigma_{\pi}^{(g_{A}^{2})}(\vec{k})=\cdots+\tfrac{2g_{A}^{2}\alpha_{t}^{2}%
}{F_{\pi}^{2}}e^{-6h}Y\left[  \sin^{2}(k_{\ast1})+\sin^{2}(k_{\ast2})+\sin
^{2}(k_{\ast3})\right]  . \label{improvepi}%
\end{equation}

\section{Two-loop average energy}

\begin{figure}
[ptb]
\begin{center}
\includegraphics[
height=1.772in,
width=1.5342in
]%
{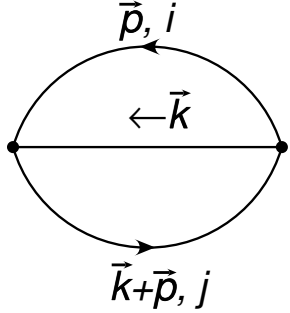}%
\caption{Two loop connected bubble diagram at $O(g_{A}^{2})$.}%
\label{football}%
\end{center}
\end{figure}

We will calculate the shift in the average energy by computing%
\begin{equation}
-\frac{\partial}{\partial\beta}\ln\left[  \frac{Z(g_{A}^{2},C)}{Z(0,0)}%
\right]  .
\end{equation}
The logarithm of the full partition function is the sum of the connected
diagrams. \ At $O(g_{A}^{2})$ we get a contribution from the connected bubble
diagram shown in Fig. \ref{football}.%

\begin{figure}
[ptbptb]
\begin{center}
\includegraphics[
height=2.0358in,
width=0.9055in
]%
{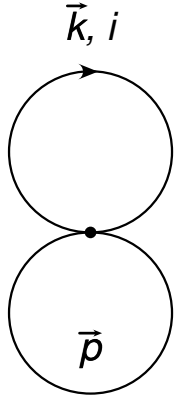}%
\caption{Two loop connected bubble due to a $\pi\pi\bar{N}N$ interaction in
the temporally-improved action at $O(g_{A}^{2})$.}%
\label{earringpi}%
\end{center}
\end{figure}

The amplitude for this bubble can be obtained in a straightforward manner from
either (\ref{neutronselfg2}) or (\ref{piselfg2})$:$%
\begin{equation}
-\tfrac{g_{A}^{2}\alpha_{t}^{2}e^{-12h}}{F_{\pi}^{2}L^{3}L_{t}}\sum_{\vec
{p},\vec{k}}D_{N}(\vec{p})D_{N}(\vec{k}+\vec{p})D_{\pi}(\vec{k})\left[
\sin^{2}(k_{\ast1})+\sin^{2}(k_{\ast2})+\sin^{2}(k_{\ast3})\right]  .
\label{delta_g2}%
\end{equation}
In our temporally-improved action the term%
\begin{equation}
S=...-\tfrac{g_{A}^{2}\alpha_{t}^{2}}{8F_{\pi}^{2}}e^{-6h}\sum_{\vec{n}%
,\hat{l}_{s},i}c_{i}^{\ast}(\vec{n})c_{i}^{\prime}(\vec{n})(\Delta_{l_{s}%
}^{\pm}\pi_{0}(n))^{2}.
\end{equation}
produces the diagram shown in Fig. \ref{earringpi}.%
The amplitude for this process can be computed from either
(\ref{improveneutron}) or (\ref{improvepi}) and is
\begin{equation}
\tfrac{g_{A}^{2}\alpha_{t}^{2}}{F_{\pi}^{2}}e^{-6h}Y\sum_{\vec{k}}D_{\pi}%
(\vec{k})\left[  \sin^{2}(k_{\ast1})+\sin^{2}(k_{\ast2})+\sin^{2}(k_{\ast
3})\right]  =\tfrac{g_{A}^{2}\alpha_{t}^{2}L^{3}L_{t}}{4F_{\pi}^{2}}e^{-6h}XY.
\label{improvedelta}%
\end{equation}

\begin{figure}
[ptb]
\begin{center}
\includegraphics[
height=2.0349in,
width=0.8864in
]%
{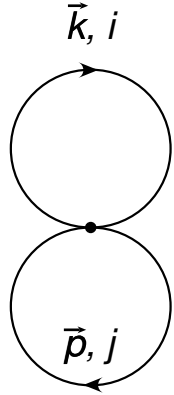}%
\caption{Two loop connected bubble at $O(C)$.}%
\label{earrings}%
\end{center}
\end{figure}

At $O(C)$ we have the\ connected bubble diagram shown in Fig. \ref{earrings}.%
If the vertex is located at lattice site $\vec{n}$ the lowest order
interaction is
\begin{equation}
e^{-12h}\left(  e^{-C\alpha_{t}}-1\right)  c_{\uparrow}^{\ast}(\vec
{n})c_{\uparrow}^{\prime}(\vec{n})c_{\downarrow}^{\ast}(\vec{n})c_{\downarrow
}^{\prime}(\vec{n}).
\end{equation}
Summing over all sites, we find that the connected bubble in Fig.
\ref{earrings} has the amplitude%
\begin{equation}
L^{3}L_{t}e^{-12h}(e^{-C\alpha_{t}}-1)Y^{2}. \label{twoloop_c}%
\end{equation}

\section{Average energy from simulations}

The average energy can be computed by taking $-\frac{\partial}{\partial\beta
}\ln Z$ and then adding $\mu A$ to the result, where $A$ is the average number
of nucleons. \ The partition function is given by%
\begin{equation}
Z(\beta)\propto e^{-2(m_{N}-\mu)\beta L^{3}}e^{\tfrac{C\beta}{2}L^{3}}\int
D\pi DsDc^{\prime}Dc^{\ast}\exp\left[  -S_{\bar{N}N+\pi\bar{N}N+s\bar{N}%
N}-S_{\pi\pi}-S_{ss}\right]  .
\end{equation}
It is convenient to define a new partition function $Z^{\prime}$ with
normalization,%
\begin{equation}
Z^{\prime}(\beta)=\int D\pi DsDc^{\prime}Dc^{\ast}\exp\left[  -S_{\bar{N}%
N+\pi\bar{N}N+s\bar{N}N}-S_{\pi\pi}-S_{ss}\right]  .
\end{equation}
$Z^{\prime}$ is what we actually compute in the simulation. \ Then%
\begin{equation}
-\frac{\partial}{\partial\beta}\ln Z=-\frac{\partial}{\partial\beta}\ln
Z^{\prime}+2(m_{N}-\mu)L^{3}-\tfrac{C}{2}L^{3}.
\end{equation}

After computing $-\frac{\partial}{\partial\beta}\ln Z$ we subtract out the
value for $-\frac{\partial}{\partial\beta}\ln Z$ at the same $\beta$ but zero
neutron density. \ In our theory we have not included pion self-interactions.
\ Therefore in the absence of neutrons we can calculate $-\frac{\partial
}{\partial\beta}\ln Z$ for free pions exactly. \ If we later decide to include
pion self-interactions then a separate simulation with only pions will be
needed for this calculation.

At zero neutron density and free neutral pions, the lattice path integral
gives%
\begin{equation}
Z^{\prime}\propto\left[  \det\left(  \left(  S_{\pi\pi}\right)  _{ij}\right)
\right]  ^{-1/2}%
\end{equation}
where $\left(  S_{\pi\pi}\right)  _{ij}$ are the coefficients of the quadratic
form in the pion action $S_{\pi\pi}$. \ We find%
\begin{align}
&  -\frac{\partial}{\partial\beta}\ln\left[  \left[  \det\left(  \left(
S_{\pi\pi}\right)  _{ij}\right)  \right]  ^{-1/2}\right] \nonumber\\
&  =\tfrac{1}{2}\frac{\partial}{\partial\beta}\sum_{\vec{k}}\ln\left[
(\tfrac{m_{\pi}^{2}}{2}+3)\alpha_{t}+\alpha_{t}^{-1}-\sum_{\hat{l}}e_{l}%
\cos(k_{\ast l})\right] . \label{pionenergy}%
\end{align}
This quantity is the pion contribution to $-\frac{\partial}{\partial\beta}\ln
Z^{\prime}$. \ This is not exactly what one might define as the pion
contribution to the energy, though it is closely related.\ \ To calculate the
pion contribution to the energy, one also needs to add%
\begin{equation}
\tfrac{1}{2}L^{3}L_{t}\frac{\partial}{\partial\beta}\ln(\alpha_{t})=\tfrac
{1}{2}L^{3}L_{t}\beta^{-1},
\end{equation}
which arises from an additional factor of%
\begin{equation}
\left[  \tfrac{1}{\sqrt{\alpha_{t}}}\right]  ^{L^{3}L_{t}}%
\end{equation}
in the partition function. \ This factor is due to the conjugate momenta
integrations going from the Hamiltonian formalism to the path integral
formalism,%
\begin{align}
\int dp\left\langle q_{n+1}\right\vert \exp\left[  \tfrac{\Delta T}{2}%
\tfrac{\partial^{2}}{\partial^{2}q}\right]  \left\vert p\right\rangle
\left\langle p\right\vert \left.  q_{n}\right\rangle  &  \propto\int
dp\exp\left[  -\tfrac{\Delta T}{2}p^{2}+ip(q_{n+1}-q_{n})\right] \nonumber\\
&  \propto\tfrac{1}{\sqrt{\Delta T}}\exp\left[  -\tfrac{1}{2\Delta T}%
(q_{n+1}-q_{n})^{2}\right]  .
\end{align}

After we calculate $-\frac{\partial}{\partial\beta}\ln Z$, we can compute the
average energy per nucleon using%
\begin{equation}
\frac{E}{A}=-\frac{1}{A}\frac{\partial}{\partial\beta}\ln Z+\mu\text{.}%
\end{equation}
In the simulations $L_{t}$ is kept fixed, and $\alpha_{t}$ is varied in order
to compute $\frac{\partial}{\partial\beta}$.

\section{Weak coupling results}

In this section we check that the results of our numerical simulations at weak
coupling agree with our results from perturbation theory. \ For the neutrons
we define the temporal two-point correlation functions,
\begin{equation}
\left\langle a%
\genfrac{}{}{0pt}{1}{n_{t}}{\longleftrightarrow}%
a^{\dagger}\right\rangle \equiv Z^{-1}\text{Tr}\left[  \exp(-(\beta
-n_{t}\alpha_{t})H)a_{\uparrow}(n_{t},0,0,0)\exp(-n_{t}\alpha_{t}%
H)a_{\uparrow}^{\dagger}(0,0,0,0)\right]
\end{equation}
and spatially separated correlation functions in the $x$ direction%
\begin{equation}
\left\langle a%
\genfrac{}{}{0pt}{1}{n_{s}}{\longleftrightarrow}%
a^{\dagger}\right\rangle \equiv Z^{-1}\text{Tr}\left[  \exp(-\beta
H)a_{\uparrow}(0,n_{s},0,0)a_{\uparrow}^{\dagger}(0,0,0,0)\right]  .
\end{equation}
The results are exactly the same in the $y$ and $z$ directions. \ Similarly
for the neutral pion, we define the temporal and spatial correlation functions%
\begin{equation}
\left\langle \pi%
\genfrac{}{}{0pt}{1}{n_{t}}{\longleftrightarrow}%
\pi\right\rangle \equiv Z^{-1}\text{Tr}\left[  \exp(-(\beta-n_{t}\alpha
_{t})H)\pi(n_{t},0,0,0)\exp(-n_{t}\alpha_{t}H)\pi(0,0,0,0)\right]  ,
\end{equation}%
\begin{equation}
\left\langle \pi%
\genfrac{}{}{0pt}{1}{n_{s}}{\longleftrightarrow}%
\pi\right\rangle \equiv Z^{-1}\text{Tr}\left[  \exp(-\beta H)\pi
(0,n_{s},0,0)\pi(0,0,0,0)\right]  .
\end{equation}

At weak coupling we compare the temporal and spatial correlation functions for
the neutron and pion as well as the energy per neutron, $E/A$. $\ $We use the
parameters $a^{-1}=150$ MeV, $\beta=2.0$, $L=3$, $L_{t}=3$, $m_{N}^{phys}=939$
MeV$,$ and $m_{\pi}^{phys}=135$ MeV, $\mu=m_{N}-0.1$. $\ $We first take
$g_{A}=0$ and $C=-0.135$. \ In Tables 1a and 1b we show the results for the
free neutron correlation functions, the one-loop results using the self-energy
correction given in (\ref{self_c}), and the results of our Monte Carlo simulation.%

\[%
\genfrac{}{}{0pt}{}{\text{Table 1a: \ }\left\langle
a\genfrac{}{}{0pt}{1}{n_{t}}{\longleftrightarrow}a^{\dagger}\right\rangle
\text{ for }g_{A}=0,C=-0.135}{%
\begin{tabular}
[c]{|l|l|l|l|}\hline
$n_{t}$ & $0$ & $1$ & $2$\\\hline
Free & $0.7568$ & $0.5027$ & $0.3444$\\\hline
One-loop & $0.7453$ & $0.5059$ & $0.3537$\\\hline
Simulation & $0.7447(2)$ & $0.5057(3)$ & $0.3537(3)$\\\hline
\end{tabular}
}%
\]

\[%
\genfrac{}{}{0pt}{}{\text{Table 1b: \ }\left\langle
a\genfrac{}{}{0pt}{1}{n_{s}}{\longleftrightarrow}a^{\dagger}\right\rangle
\text{ for }g_{A}=0,C=-0.135}{%
\begin{tabular}
[c]{|l|l|}\hline
$n_{s}$ & $1$\\\hline
Free & $-0.03903$\\\hline
One-loop & $-0.03940$\\\hline
Simulation & $-0.03936(2)$\\\hline
\end{tabular}
}%
\]
In Table 1c we show the free neutron value for the energy per neutron, the
two-loop corrected value using (\ref{twoloop_c}), and the result of the simulation.%

\[%
\genfrac{}{}{0pt}{}{\text{Table 1c: \ }E/A\text{ for }g_{A}=0,C=-0.135}{%
\begin{tabular}
[c]{|l|l|}\hline
Free & $6.665$\\\hline
One-loop & $6.653$\\\hline
Simulation & $6.652(1)$\\\hline
\end{tabular}
}%
\]
There is no correction to the free pion correlation function when $g_{A}=0$.
\ We see that all the simulation results match the loop calculations for
$g_{A}=0$ and small $C$.

For $C=0$ and small $g_{A}$ there are two sets of diagrams which we would like
to separately compare with simulation results. $\ $The temporally-improved
neutron action has the form%
\begin{align}
S_{\bar{N}N+\pi\bar{N}N}  &  =\sum_{\vec{n},i}e^{(m_{N}-\mu)\alpha_{t}}%
c_{i}^{\ast}(\vec{n})c_{i}^{\prime}(\vec{n}+\hat{0})\nonumber\\
&  -\sum_{\vec{n},i,j}c_{i}^{\ast}(\vec{n})\left(  e^{-X(\vec{n})}\right)
_{ij}c_{j}^{\prime}(\vec{n})\nonumber\\
&  -h\sum_{\vec{n},\hat{l}_{s},i}\left[  c_{i}^{\ast}(\vec{n})c_{i}(\vec
{n}+\hat{l}_{s})+c_{i}^{\ast}(\vec{n})c_{i}(\vec{n}-\hat{l}_{s})\right]
\end{align}
where $X_{ij}(\vec{n})$%
\begin{equation}
\left[
\begin{array}
[c]{cc}%
6h-\tfrac{g_{A}\alpha_{t}}{2F_{\pi}}\Delta_{3}^{\pm}\pi_{0}(\vec{n}) &
-\tfrac{g_{A}\alpha_{t}}{2F_{\pi}}\left(  \Delta_{1}^{\pm}-i\Delta_{2}^{\pm
}\right)  \pi_{0}(\vec{n})\\
-\tfrac{g_{A}\alpha_{t}}{2F_{\pi}}\left(  \Delta_{1}^{\pm}+i\Delta_{2}^{\pm
}\right)  \pi_{0}(\vec{n}) & 6h+\tfrac{g_{A}\alpha_{t}}{2F_{\pi}}\Delta
_{3}^{\pm}\pi_{0}(\vec{n})
\end{array}
\right]  .
\end{equation}
The one-loop corrected neutron correlator gets a contribution from both
(\ref{neutronselfg2}) and (\ref{improveneutron}); the one-loop corrected pion
correlator uses (\ref{piselfg2}) and (\ref{improvepi}); and the one-loop
corrected energy per neutron has terms (\ref{delta_g2}) and
(\ref{improvedelta})$.$ \ The comparisons with simulation results for
$g_{A}=0.750$, $C=0$, are shown in Tables 2a-e and are labelled by `exp',
which stands for the exponential form used in the temporally-improved action.

We will also remove the temporally-improved diagrams which gave us the
contributions (\ref{improveneutron}), (\ref{improvepi}), and
(\ref{improvedelta}). \ We do this by replacing the term in the action%
\begin{equation}
-\sum_{\vec{n},i,j}c_{i}^{\ast}(\vec{n})\left(  e^{-X(\vec{n})}\right)
_{ij}c_{j}^{\prime}(\vec{n})
\end{equation}
by%
\begin{equation}
\sum_{\vec{n},i,j}c_{i}^{\ast}(\vec{n})M_{ij}(\vec{n})c_{j}^{\prime}(\vec{n}),
\end{equation}
where $M_{ij}(\vec{n})$ is%
\begin{equation}
\exp(-6h)\left[
\begin{array}
[c]{cc}%
-1-\tfrac{g_{A}\alpha_{t}}{2F_{\pi}}\Delta_{3}^{\pm}\pi_{0}(\vec{n}) &
-\tfrac{g_{A}\alpha_{t}}{2F_{\pi}}\left(  \Delta_{1}^{\pm}-i\Delta_{2}^{\pm
}\right)  \pi_{0}(\vec{n})\\
-\tfrac{g_{A}\alpha_{t}}{2F_{\pi}}\left(  \Delta_{1}^{\pm}+i\Delta_{2}^{\pm
}\right)  \pi_{0}(\vec{n}) & -1+\tfrac{g_{A}\alpha_{t}}{2F_{\pi}}\Delta
_{3}^{\pm}\pi_{0}(\vec{n})
\end{array}
\right]  .
\end{equation}
The comparisons with simulation results for this linearized action are also
shown in Tables 2a-e and are labelled by `lin'.%

\[%
\genfrac{}{}{0pt}{}{\text{Table 2a: \ }\left\langle
a\genfrac{}{}{0pt}{1}{n_{t}}{\longleftrightarrow}a^{\dagger}\right\rangle
\text{ for }g_{A}=0.750,C=0}{%
\begin{tabular}
[c]{|l|l|l|l|}\hline
$n_{t}$ & $0$ & $1$ & $2$\\\hline
Free & $0.7568$ & $0.5027$ & $0.3444$\\\hline
One-loop (lin) & $0.7586$ & $0.4978$ & $0.3400$\\\hline
Simulation (lin) & $0.7585(1)$ & $0.4974(1)$ & $0.3399(1)$\\\hline
One-loop (exp) & $0.7496$ & $0.5005$ & $0.3475$\\\hline
Simulation (exp) & $0.7494(1)$ & $0.5000(1)$ & $0.3472(1)$\\\hline
\end{tabular}
}%
\]

\[%
\genfrac{}{}{0pt}{}{\text{Table 2b: \ }\left\langle
a\genfrac{}{}{0pt}{1}{n_{s}}{\longleftrightarrow}a^{\dagger}\right\rangle
\text{ for }g_{A}=0.750,C=0}{%
\begin{tabular}
[c]{|l|l|}\hline
$n_{s}$ & $1$\\\hline
Free & $-0.03903$\\\hline
One-loop (lin) & $-0.03859$\\\hline
Simulation (lin) & $-0.03856(1)$\\\hline
One-loop (exp) & $-0.03890$\\\hline
Simulation (exp) & $-0.03886(1)$\\\hline
\end{tabular}
}%
\]%
\[%
\genfrac{}{}{0pt}{}{\text{Table 2c: \ }\left\langle \pi
\genfrac{}{}{0pt}{1}{n_{t}}{\longleftrightarrow}\pi\right\rangle \text{ for
}g_{A}=0.750,C=0}{%
\begin{tabular}
[c]{|l|l|l|}\hline
$n_{t}$ & $0$ & $1$\\\hline
Free & $0.1764$ & $0.0615$\\\hline
One-loop (lin) & $0.1780$ & $0.0644$\\\hline
Simulation (lin) & $0.1780(3)$ & $0.0643(3)$\\\hline
One-loop (exp) & $0.1810$ & $0.0660$\\\hline
Simulation (exp) & $0.1809(2)$ & $0.0659(2)$\\\hline
\end{tabular}
}%
\]%
\[%
\genfrac{}{}{0pt}{}{\text{Table 2d: \ }\left\langle \pi
\genfrac{}{}{0pt}{1}{n_{s}}{\longleftrightarrow}\pi\right\rangle \text{ for
}g_{A}=0.750,C=0}{%
\begin{tabular}
[c]{|l|l|}\hline
$n_{s}$ & $1$\\\hline
Free & $0.0364$\\\hline
One-loop (lin) & $0.0366$\\\hline
Simulation (lin) & $0.0364(2)$\\\hline
One-loop (exp) & $0.0367$\\\hline
Simulation (exp) & $0.0365(2)$\\\hline
\end{tabular}
}%
\]%
\[%
\genfrac{}{}{0pt}{}{\text{Table 2e: \ }E/A\text{ for }g_{A}=0.750,C=0}{%
\begin{tabular}
[c]{|l|l|}\hline
Free & $6.665$\\\hline
One-loop (lin) & $6.673$\\\hline
Simulation (lin) & $6.672(1)$\\\hline
One-loop (exp) & $6.640$\\\hline
Simulation (exp) & $6.638(1)$\\\hline
\end{tabular}
}%
\]
We see that all the simulation results match the loop calculations for $C=0$
and small $g_{A}$.

\section{Renormalization of coefficients}

We now discuss the renormalization of operator coefficients in our lowest
order effective Lagrangian. \ At zero temperature and $\mu<m_{N}$, the pion
self-energy vanishes since there are no neutron holes. \ Thus there is no
renormalization for the pion wavefunction and mass.

In the Weinberg counting scheme, the neutron self-energy at zero temperature
and $\mu<m_{N}$ gets a contribution from diagrams such as the one shown in
Fig. \ref{n_self_n_pi}. \ This is the lowest order diagram and comes at chiral
order $\nu=3$. \ Since we require cutoff independence, the counterterm
diagrams must also be at order $\nu=3$. \ Since this is a small correction, we
will ignore wavefunction and kinetic energy renormalization for the neutron in
the present study. \ Although the mass counterterm is also small, we will take
some extra care with this one since we are interested in precise measurements
of the energy per neutron. \ In the non-relativistic formalism the mass
counterterm can be regarded as a shift in the definition of the chemical
potential. \ Its purpose is to eliminate cutoff dependence in loop diagrams,
but we will also use it to absorb residual effects due to the finite temporal
spacing,
\begin{equation}
\alpha_{t}=\tfrac{a_{t}}{a}>0\text{.}%
\end{equation}
We will refer to the mass counterterm as $\Delta m_{N}.$ \ In the limit of
zero neutron density, the neutrons behave as free particles. \ We can
therefore calculate $\Delta m_{N}$ in that limit. \ From a theoretical point
of view it would be nice to measure the average energy at both zero density
and zero temperature. \ Computationally, however, it is more practical to make
the measurement at non-zero temperature.

\begin{figure}
[ptb]
\begin{center}
\includegraphics[
height=1.1623in,
width=1.5342in
]%
{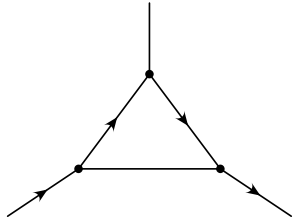}%
\caption{One-loop correction to the one-particle irreducible $\pi\bar{N}N$
vertex.}%
\label{vertex_pi_n}%
\end{center}
\end{figure}

At zero temperature and $\mu<m_{N}$, the one-loop contribution to the
one-particle irreducible $\pi\bar{N}N$ vertex is shown in Fig.
\ref{vertex_pi_n}.%
This process is also at chiral order $\nu=3$. \ Since this is a small
correction we will also ignore renormalization of the $\pi\bar{N}N$
coefficient and set its value to equal the physically measured value,
\begin{equation}
g_{A}\approx1.25.
\end{equation}

\begin{figure}
[ptb]
\begin{center}
\includegraphics[
height=1.5965in,
width=1.5342in
]%
{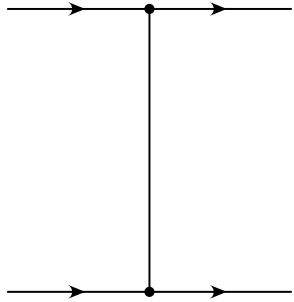}%
\caption{Two-particle irreducible one pion exchange diagram.}%
\label{twopi_ladder}%
\end{center}
\end{figure}

\begin{figure}
[ptbptb]
\begin{center}
\includegraphics[
height=1.5342in,
width=1.5342in
]%
{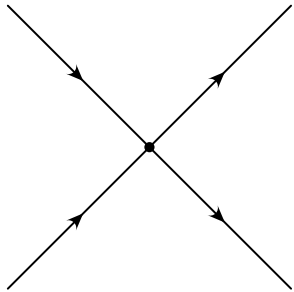}%
\caption{Two-particle irreducible contact diagram.}%
\label{twopi_contact}%
\end{center}
\end{figure}

At zero temperature and $\mu<m_{N}$, the lowest order contribution to the $NN$
scattering Green's function is due to iterating the lowest order two-particle
irreducible diagrams. \ The lowest order two-particle irreducible diagrams are
shown in Figs. \ref{twopi_ladder} and \ref{twopi_contact}.%
When there is a bound state or the scattering length is very large compared to
other relevant scales, the two-particle irreducible kernel must be iterated
and summed to all orders. \ The result is that the $NN$ scattering Green's
function has a cutoff dependence that cannot be regarded as a small
correction. \ We will therefore need a non-perturbative calculation of the
$\bar{N}N\bar{N}N$ contact interaction counterterm. \ We will use the
Schr\"{o}dinger equation on the lattice to deal with this problem, and we
describe the procedure in the next section.

\section{Lattice Schr\"{o}dinger equation and phase shifts}

We adjust $C$, the coefficient of the $\bar{N}N\bar{N}N$ contact interaction,
so that the $NN$ s-wave scattering length matches the experimental value (see
for example \cite{Funk:2000mf}). \ In order to calculate the phase shifts, we
will solve the lattice Schr\"{o}dinger equation for the two-neutron system and
observe the asymptotic form of the scattering wavefunctions. \ The first step
will be to construct the potential between two neutrons.

Let $\left\vert 0\right\rangle $ be the free non-interacting vacuum. \ The
two-neutron state with zero total spatial momentum, zero total intrinsic spin,
and spatial separation $\vec{n}_{s}$ can be constructed as%

\begin{equation}
\left\vert \vec{n}_{s}\right\rangle =\tfrac{1}{\sqrt{2L^{3}}}\sum_{m}\left[
a_{\uparrow}^{\dag}(\vec{n}_{s}+\vec{m}_{s})a_{\downarrow}^{\dag}(\vec{m}%
_{s})-a_{\downarrow}^{\dag}(\vec{n}_{s}+\vec{m}_{s})a_{\uparrow}^{\dag}%
(\vec{m}_{s})\right]  \left\vert 0\right\rangle .
\end{equation}
We let $V_{2N}$ be the lowest-order potential in the Weinberg counting scheme
between two-neutron states with zero total spatial momentum, zero total
intrinsic spin, and spatial separation $\vec{n}_{s}$. \ Using the fact that%
\begin{equation}
\sum_{\vec{n}_{s}^{\prime},\vec{n}_{s}^{\prime\prime}}\left\langle \vec{n}%
_{s}\right\vert :a_{i^{\prime\prime}}^{\dag}(\vec{n}_{s}^{\prime\prime}%
)\sigma_{i^{\prime\prime}j^{\prime\prime}}^{l_{s}^{\prime\prime}}%
a_{j^{\prime\prime}}(\vec{n}_{s}^{\prime\prime})a_{i^{\prime}}^{\dag}(\vec
{n}_{s}^{\prime})\sigma_{i^{\prime}j^{\prime}}^{l_{s}^{\prime}}a_{j^{\prime}%
}(\vec{n}_{s}^{\prime}):\left\vert \vec{n}_{s}\right\rangle =-2\delta
_{l_{s}^{\prime},l_{s}^{\prime\prime}}\left\langle \vec{n}_{s}\right.
\left\vert \vec{n}_{s}\right\rangle ,
\end{equation}
we have%
\begin{equation}
V_{2N}(\vec{n}_{s})=\frac{g_{A}^{2}}{2F_{\pi}^{2}L^{3}}\sum_{\vec{k}_{s}}%
\frac{e^{-i\vec{n}_{s}\cdot\vec{k}_{s\ast}}\sum_{l_{s}}\sin^{2}(k_{s\ast
})_{l_{s}}}{\tfrac{m_{\pi}^{2}}{2}+3-\sum_{l_{s}}\cos(k_{s\ast})_{l_{s}}%
}+C\delta_{\vec{n}_{s},0}. \label{potential2}%
\end{equation}

\begin{figure}
[ptb]
\begin{center}
\includegraphics[
height=2.7518in,
width=3.7723in
]%
{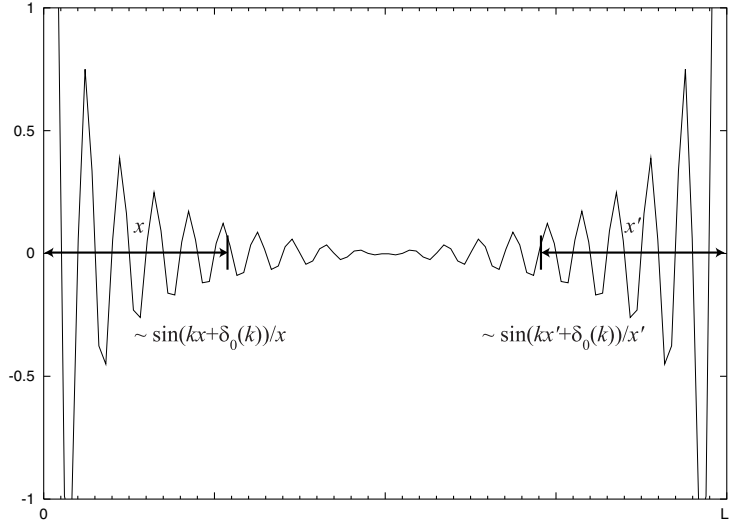}%
\caption{Measuring s-wave phase shifts from the asymptotic form of scattering
states in a periodic box of length $L$. The center of the potential is at $x =
0$ and $x = L$}%
\label{asymptoticbessel}%
\end{center}
\end{figure}

After obtaining $V_{2N}$ we can construct a matrix representation for the
Hamiltonian in the two-neutron sector and solve the time-independent lattice
Schr\"{o}dinger equation. \ At this stage one could implement L\"{u}scher's
formula for the measuring phase shifts in a cubical periodic box
\cite{Luscher:1991ux}\cite{Luscher:1986pf}. \ However in our case we can
construct the eigenvectors explicitly using Lanczos iteration, and so we find
it more straightforward and accurate to read the phaseshifts directly from the
asymptotic forms of the s-wave scattering states. \ Since we are working in a
periodic box it is important to measure the phase shifts far away from the
center of the potential and all its translations in the periodic box, as shown
in Fig. \ref{asymptoticbessel}.%

Before using this technique for our actual neutron system, we first test our
technique for hard sphere scattering where the exact result for the s-wave
phase shifts is well known. \ If the spheres have radii $r/2$ (therefore the
centers of the spheres are separated by $r$) then the s-wave phase shift has
the form%
\begin{equation}
\delta_{0}=-kr,
\end{equation}
where $k$ is the momentum. \ The momentum $k$ for a given scattering state can
be determined from the energy and the free particle non-relativistic
dispersion relation. \ In Table 3 we show results for the s-wave phase shift
$\delta_{0}$ at inverse lattice spacing $a^{-1}=150$ MeV, with particle masses
set at $m_{N}$ and a lattice volume of $20^{3}$. \ We use several radii $r$
and show comparisons with the exact continuum result $-kr$. \ The results
suggest that our lattice Schr\"{o}dinger technique seems to be functioning properly.%

\[%
\genfrac{}{}{0pt}{}{\text{Table 3: \ Hard sphere scattering phase shifts}}{%
\begin{tabular}
[c]{|l|l|l|l|}\hline
$r$ & $k$ & $\delta_{0}$ & $-kr$\\\hline
$4.5$ & $0.117$ & $-0.56$ & $-0.53$\\\hline
$5.5$ & $0.147$ & $-0.85$ & $-0.81$\\\hline
$6.5$ & $0.179$ & $-1.20$ & $-1.16$\\\hline
\end{tabular}
}%
\]

We now use the lattice Schr\"{o}dinger technique to tune the coupling $C$ to
reproduce the large scattering length that is observed in nature. \ In Table 4
we show the best fit values for $C^{phys}$ for several different lattice
spacings.%
\[%
\genfrac{}{}{0pt}{}{\text{Table 4: \ }C^{phys}\text{ for different lattice
spacings}}{%
\begin{tabular}
[c]{|l|l|}\hline
$a^{-1}$ & $C^{phys}$\\\hline
$150$\ MeV & $-4.0\cdot10^{-5}$ MeV$^{-2}$\\\hline
$200$\ MeV & $-3.4\cdot10^{-5}$ MeV$^{-2}$\\\hline
$250$\ MeV & $-3.1\cdot10^{-5}$ MeV$^{-2}$\\\hline
$300$\ MeV & $-2.9\cdot10^{-5}$ MeV$^{-2}$\\\hline
\end{tabular}
}%
\]
We can compare this with the pionless case where the only interaction is the
contact interaction. \ In this case we need only sum bubble diagrams and that
gives the relation $C^{phys}\propto a$. \ The pionless calculation on the
lattice has been discussed in \cite{Beane:2003da}.

\section{Zone determinant method}

It can be shown that fermions at inverse temperature $\beta$ with spatial
hopping parameter $h^{\prime}$ have a localization length \cite{Lee:2003a} of%
\begin{equation}
l\sim\sqrt{\beta h^{\prime}}.
\end{equation}
This idea was used in \cite{Lee:2003mb} to generate an algorithm called the
zone determinant method to speed up the calculation of determinants using LU
decomposition in nuclear lattice simulations.

The technique is relatively simple to describe. \ Let $M$ be the neutron
matrix, in general an $n\times n$ complex matrix. \ We partition the lattice
spatially into separate zones such that the length of each zone is larger than
the localization length $l$. \ Since most neutron worldlines do not cross the
zone boundaries, they would not be affected if we set the zone boundary
hopping terms to zero. \ Hence we anticipate that the determinant of $M$ can
be approximated by the product of the submatrix determinants for each spatial zone.

Let us partition the lattice into spatial zones labelled by index $j.$ \ Let
$\{P_{j}\}$ be a complete set of matrix projection operators that project onto
the lattice sites within spatial zone $j$. \ We can write%
\begin{equation}
M=\sum_{i,j}P_{j}MP_{i}=M_{0}+M_{E},
\end{equation}
where%
\begin{align}
M_{0}  &  =\sum_{i}P_{i}MP_{i},\\
M_{E}  &  =\sum_{i\neq j}P_{j}MP_{i}.
\end{align}
If the zones can be sorted into even and odd sets so that%
\begin{equation}
P_{j}MP_{i}=0
\end{equation}
whenever $i$ is even and $j$ is odd or vice-versa, then we say that the zone
partitioning is bipartite. \ We now have%
\begin{align}
\det(M)  &  =\det(M_{0})\det(1+M_{0}^{-1}M_{E})\nonumber\\
&  =\det(M_{0})\exp(\text{Tr}(\log(1+M_{0}^{-1}M_{E})))\text{.}%
\end{align}
Using an expansion for the logarithm, we have%
\begin{equation}
\det(M)=\det(M_{0})\exp\left(  \sum_{p=1}^{\infty}\frac{(-1)^{p-1}}%
{p}\text{Tr}((M_{0}^{-1}M_{E})^{p})\right)  .
\end{equation}

Let us define
\begin{equation}
\Delta_{m}=\det(M_{0})\exp\left(  \sum_{p=1}^{m}\frac{(-1)^{p-1}}{p}%
\text{Tr}((M_{0}^{-1}M_{E})^{p})\right)  .
\end{equation}
Let $\lambda_{k}(M_{0}^{-1}M_{E})$ be the eigenvalues of $M_{0}^{-1}M_{E}$ and
$R$ be the spectral radius,%
\begin{equation}
R=\max_{k=1,...,n}(\left\vert \lambda_{k}(M_{0}^{-1}M_{E})\right\vert ).
\end{equation}
It has been shown \cite{Ipsen:2003} that for $R<1,$%
\begin{equation}
\frac{\left\vert \det(M)-\Delta_{m}\right\vert }{\left\vert \Delta
_{m}\right\vert }\leq cR^{m}e^{cR^{m}},
\end{equation}
where%
\begin{equation}
c=-n\log(1-R).
\end{equation}
The spectral radius $R$ determines the convergence of our expansion. \ $R$ can
be reduced by increasing the size of the spatial zone relative to the
localization length $l$. \ In the special case where the zone partitioning is
bipartite, we note that for any odd $p,$%
\begin{equation}
\text{Tr}((M_{0}^{-1}M_{E})^{p})=0.
\end{equation}
In that case
\begin{equation}
\Delta_{2m+1}=\Delta_{2m}.
\end{equation}

In the simulations presented in this article we use the zone determinant
method to calculate neutron matrix determinants. \ We use the second order
approximation $\Delta_{2}$ with zones of the smallest possible size, a single
spatial point ([1,1,1] in the notation of \cite{Lee:2003mb}). \ An estimate of
the approximation error is discussed along with each measurement in the
results section.

\section{Results}

We have generated simulation results for $a^{-1}=150$ MeV; $\alpha_{t}%
=\tfrac{a_{t}}{a}=1.0$; temperatures $T^{phys}=25.0$ MeV and $37.5$\ MeV; and
lattice sizes $3^{3}$, $4^{4}$, and $5^{5}$. \ Half-filling at this lattice
spacing occurs at%
\begin{equation}
\frac{\rho}{\rho_{N}}=2.64,
\end{equation}
where $\rho_{N}^{phys}$ is the normal nuclear density of about $0.17$ nucleons
per fm$^{3}$. \ The calculations were performed using the zone determinant
method using the second order approximation $\Delta_{2}$ with zones consisting
of a single spatial point. \ By calculating the exact determinants of some
generated matrix configurations we estimate the systemic error for the zone
expansion to be about $<0.1\%$ for $T^{phys}=37.5$ MeV and $<0.5\%$ for
$T^{phys}=25.0$ MeV.

We have dealt with the complex action by computing the phase as an
observable,
\begin{equation}
\left\langle O\right\rangle =\frac{\sum_{[n]}O[n]e^{-\operatorname{Re}%
S[n]}e^{-i\operatorname{Im}S[n]}}{\sum_{[n]}e^{-\operatorname{Re}%
S[n]}e^{-i\operatorname{Im}S[n]}}.
\end{equation}
For the various simulations presented here we found an average phase of about
\begin{equation}
\frac{\sum_{\lbrack n]}e^{-\operatorname{Re}S[n]}e^{-i\operatorname{Im}S[n]}%
}{\sum_{[n]}e^{-\operatorname{Re}S[n]}}\sim0.95-1.00,
\end{equation}
and so this did not present a significant computational problem.%

\begin{figure}
[ptb]
\begin{center}
\includegraphics[
height=4.5576in,
width=3.1946in,angle=-90
]%
{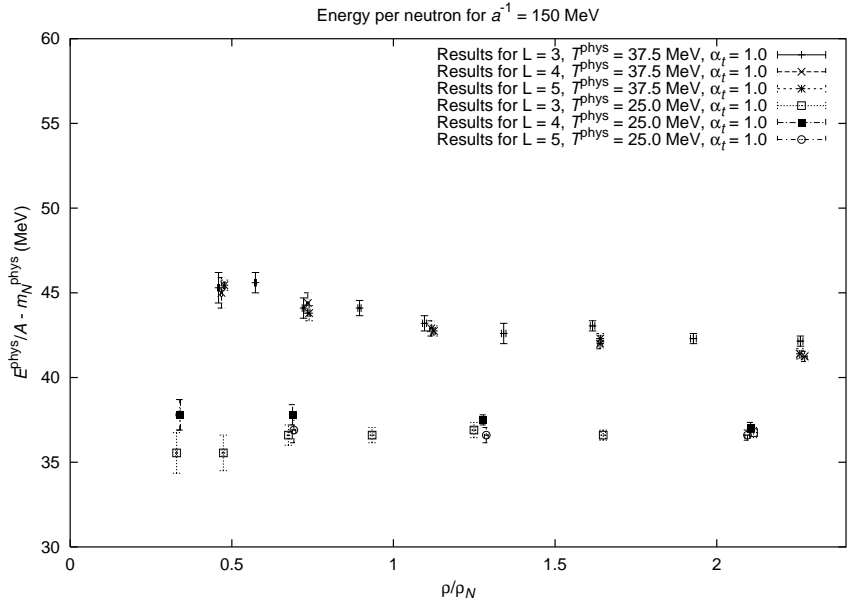}%
\caption{Energy per nucleon in MeV for temperatures 25 MeV and 37.5 MeV and
different lattice volumes.}%
\label{ldependence}%
\end{center}
\end{figure}
\begin{figure}
[ptbptb]
\begin{center}
\includegraphics[
height=4.5576in,
width=3.1946in,angle=-90
]%
{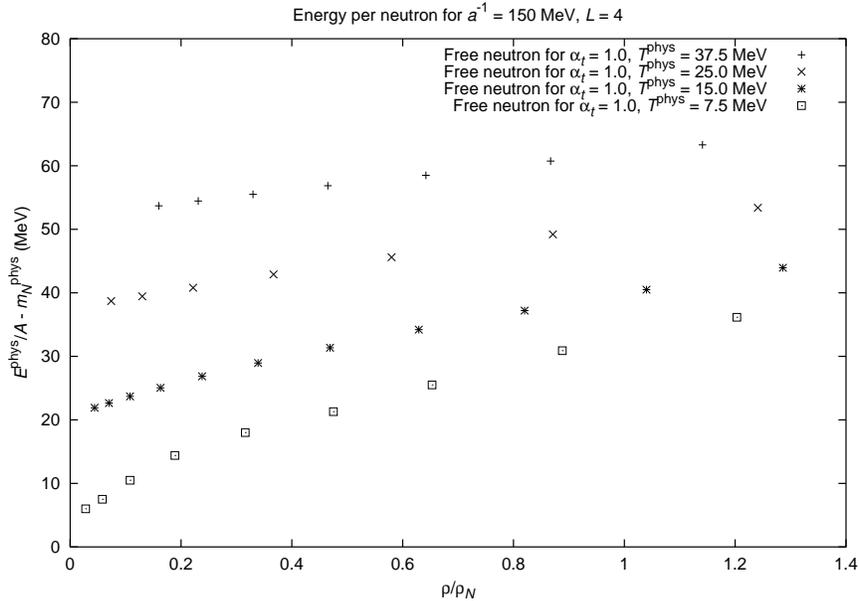}%
\caption{Energy per nucleon in MeV for temperatures 7.5, 15, 25 and 37.5 MeV
and lattice volume $4^{3}$.}%
\label{freeneutron}%
\end{center}
\end{figure}

In Fig. \ref{ldependence} we show the energy per neutron as a function of
neutron density. \ Our results indicate a rather flat function for the energy
per neutron as a function of density. \ For comparison we show in Fig.
\ref{freeneutron} the energy per neutron for the free neutron for temperatures
$T^{phys}=7.5$, $15.0$, $25.0$, and $37.5$ MeV.%

\begin{figure}
[ptb]
\begin{center}
\includegraphics[
height=4.5576in,
width=3.1946in,angle=-90
]%
{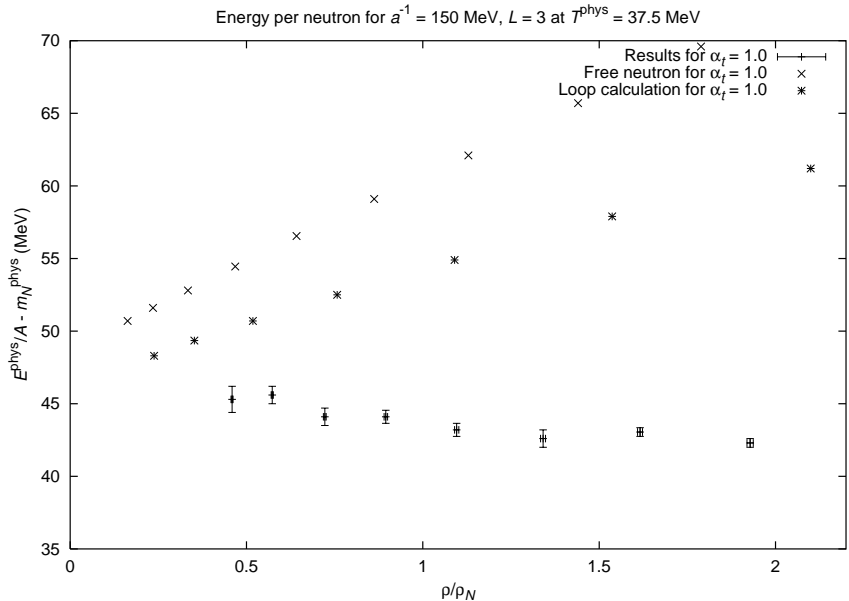}%
\caption{Energy per neutron for temperature $37.5$\ MeV and comparisons with
the free neutron result on the lattice and loop calculations.}%
\label{plotshift4}%
\end{center}
\end{figure}

\begin{figure}
[ptbptb]
\begin{center}
\includegraphics[
height=4.5576in,
width=3.1946in,angle=-90
]%
{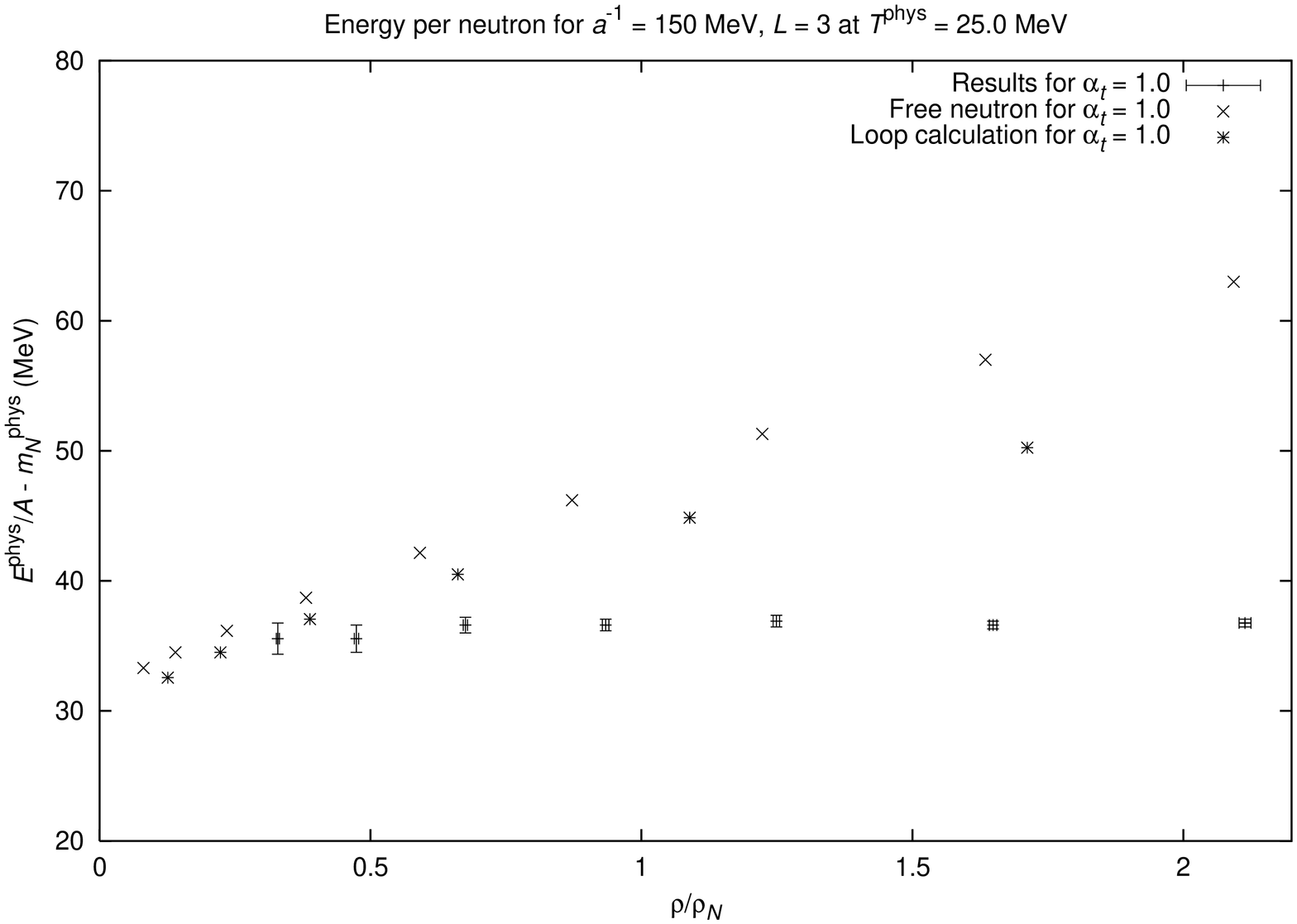}%
\caption{Energy per neutron for temperature $25.0$\ MeV and comparisons with
the free neutron result on the lattice and loop calculations.}%
\label{plotshift6}%
\end{center}
\end{figure}

In Figs. \ref{plotshift4} and \ref{plotshift6} we compare with results for
free neutrons on the lattice and loop calculations for the physical values of
$g_{A}$ and $C$. \ As is easily seen, the loop calculations are not very close
to the non-perturbative simulation results for the densities shown.%

\ \ The flatness of our energy per neutron curves at these temperatures are
intriguing and hopefully will be checked by others in the near future. \ Our
results appear to be consistent with the neutron matter results of
\cite{Muller:1999cp} at the same temperatures. Variational calculations
\cite{Friedman:1981qw} and recent quantum Monte Carlo results from
\cite{Carlson:2003wm} observe a gradually flattening of the energy per neutron
curve with increasing density, though not as flat as the results we see. \ The
calculations in \cite{Carlson:2003wm} however were performed at zero
temperature and at lower densities.

\begin{figure}
[ptb]
\begin{center}
\includegraphics[
height=4.5576in,
width=3.1946in,angle=-90
]%
{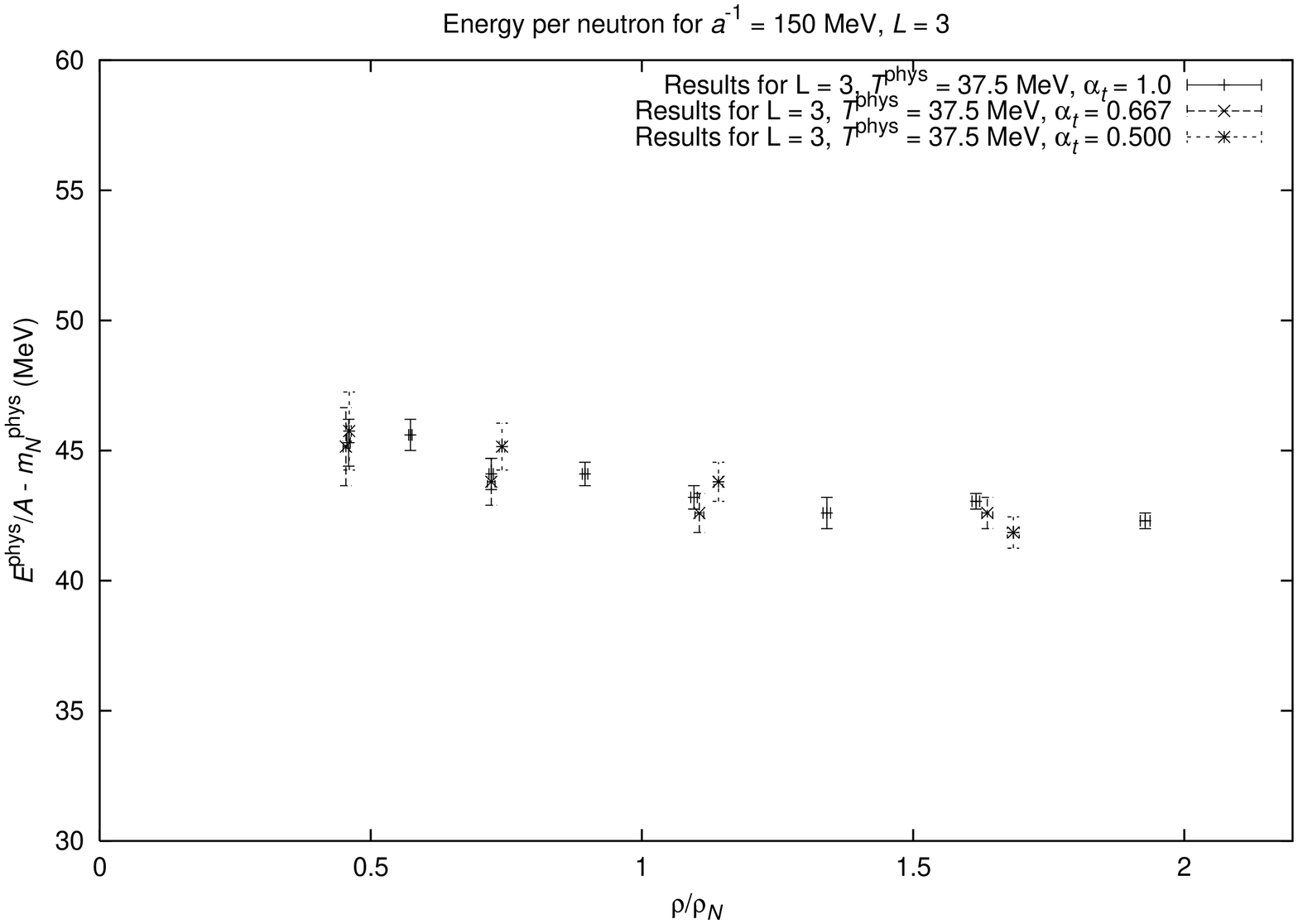}%
\caption{Dependence on $\alpha_{t}$ for temperature $37.5$ MeV.}%
\label{alphadependence}%
\end{center}
\end{figure}

We implemented a temporally-improved action in order to remove as much as
possible the dependence\ on $\alpha_{t}=\tfrac{a_{t}}{a}$, the ratio of the
temporal lattice spacing to the spatial lattice spacing. \ In Fig.
\ref{alphadependence} we show the dependence on $\alpha_{t}$ for $\alpha
_{t}=1.0$, $0.667$, $0.500$. \ We see that the dependence on $\alpha_{t}$ is
minimal.%

\begin{figure}
[ptb]
\begin{center}
\includegraphics[
height=4.5576in,
width=3.1946in,angle=-90
]%
{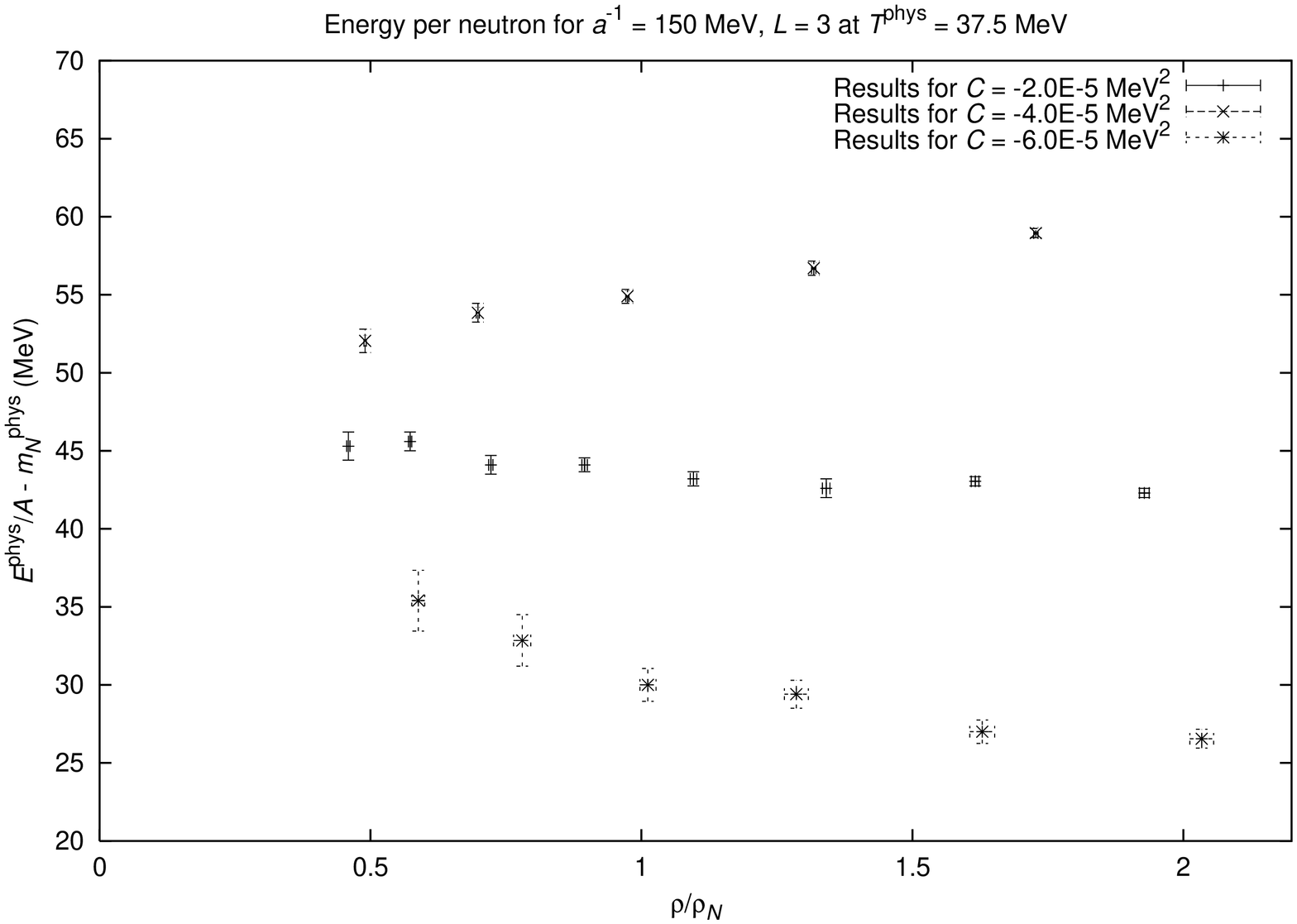}%
\caption{Dependence on $C^{phys}$ for temperature $37.5$\ MeV.}%
\label{cbeta4}%
\end{center}
\end{figure}

We now look at how the energy per neutron changes as the interaction strength
is varied. \ According to Table 4,\ the physical value for $C^{phys}$ at
$a^{-1}=150$ MeV is $-4.0\cdot10^{-5}$ MeV$^{-2}$. \ If we however take the
coupling to be $50\%$, $100\%$, and $150\%$ of the physical value we find the
results shown in Fig. \ref{cbeta4}.%

\begin{figure}
[ptbptb]
\begin{center}
\includegraphics[
height=4.5576in,
width=3.1946in,angle=-90
]%
{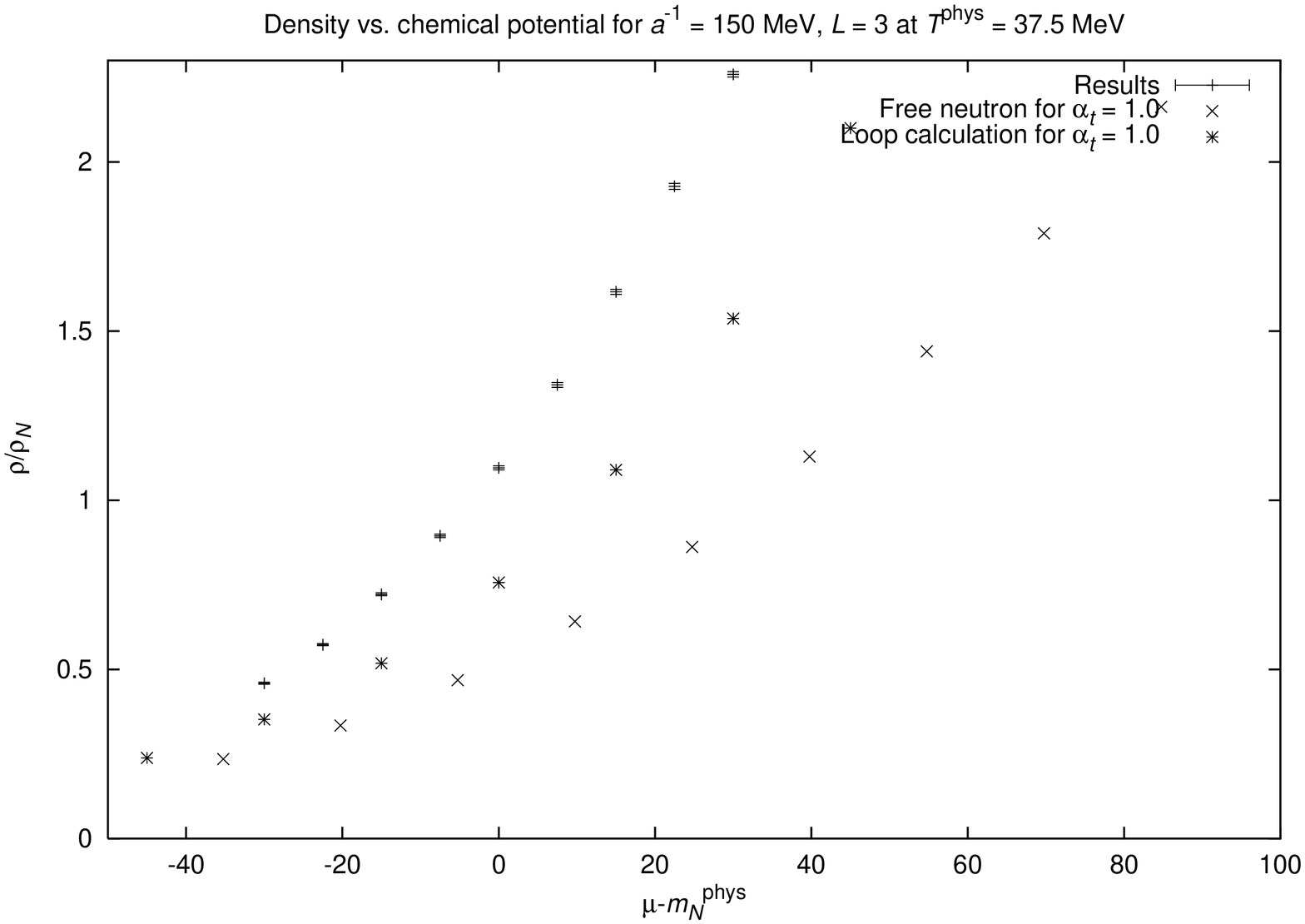}%
\caption{Density versus chemical potential for temperature $37.5$ MeV.}%
\label{chem4}%
\end{center}
\end{figure}

\begin{figure}
[ptbptbptb]
\begin{center}
\includegraphics[
height=4.5576in,
width=3.1946in,angle=-90
]%
{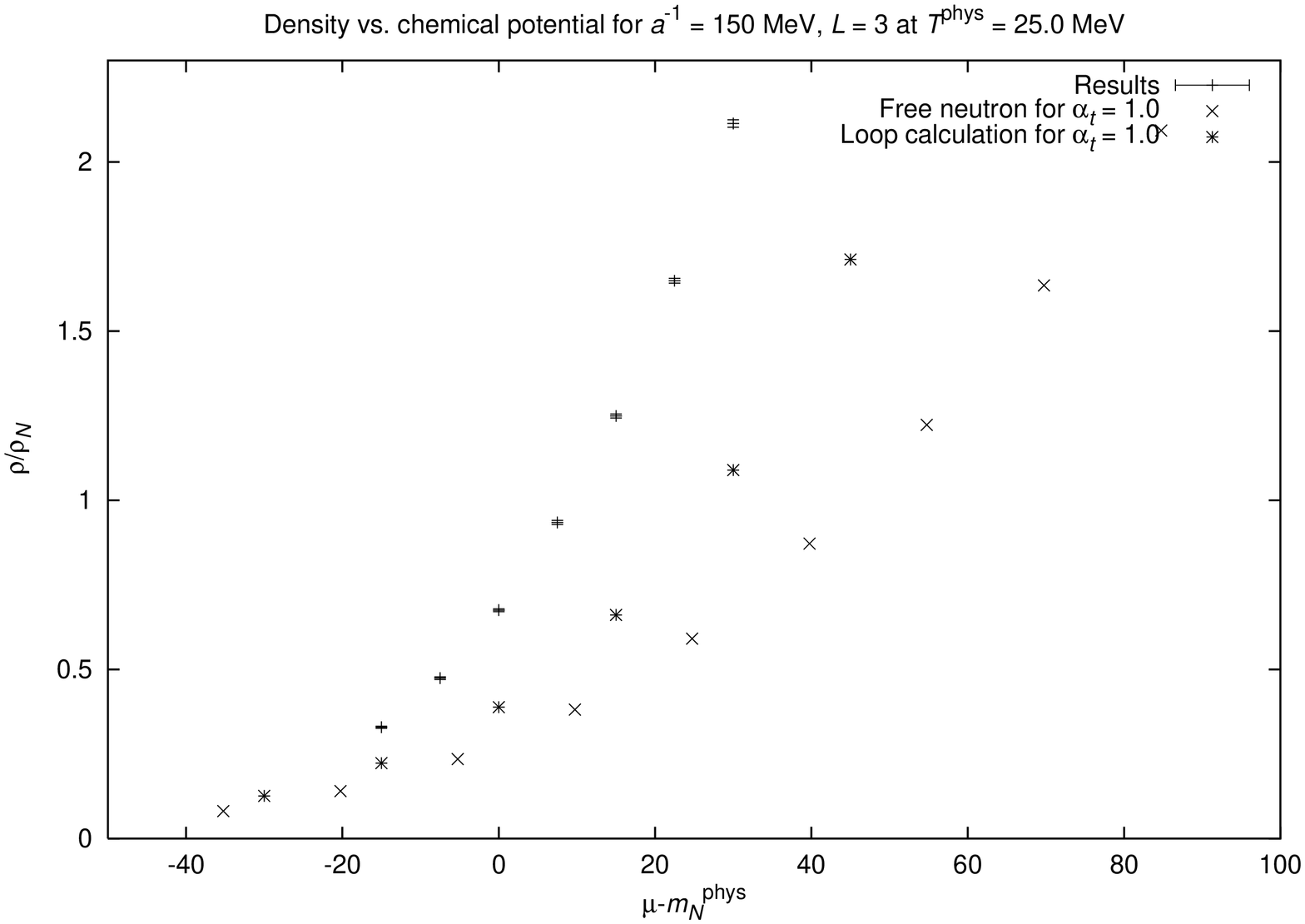}%
\caption{Density versus chemical potential for temperature $25.0$ MeV.}%
\label{chem6}%
\end{center}
\end{figure}

In Figs. \ref{chem4} and \ref{chem6} we show density versus chemical potential
and comparisons with the free neutron and loop calculations. \ We can
see\ again that loop calculations are not close to the simulation results.
\ For fixed chemical potential, the simulation density is higher than the
loop-calculated density, which is in turn higher than the free neutron
density.%

\begin{figure}
[ptb]
\begin{center}
\includegraphics[
height=4.5576in,
width=3.1946in,angle=-90
]%
{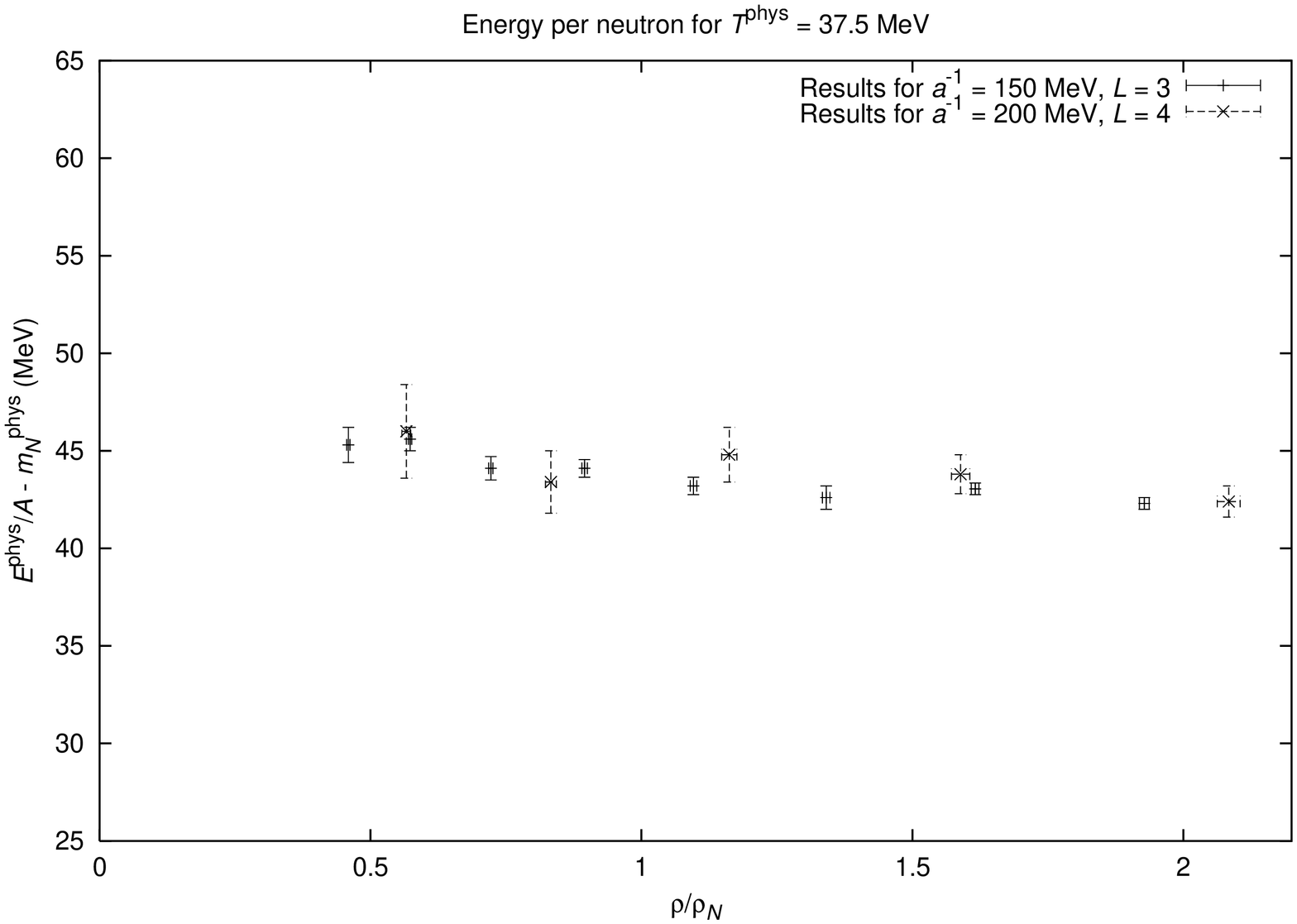}%
\caption{Comparison of the energy per neutron results for $a^{-1}=150$ MeV and
$a^{-1}=200$ MeV at temperature $37.5$ MeV.}%
\label{cutoffbeta4}%
\end{center}
\end{figure}

If our effective field theory formalism is valid, we should be able to
reproduce the same results for different lattice spacings. \ There are however
practical computational constraints on $a$. \ The determinant zone expansion
for fixed zone size and expansion order will break down if the lattice spacing
is too small. \ For $a^{-1}=200$ MeV we estimate that the second order
approximation $\Delta_{2}$ with zones consisting of a single spatial point
produces errors of size roughly $3\%$. \ In Fig. \ref{cutoffbeta4} we compare
the energy per neutron as measured for $a^{-1}=150$ MeV and $a^{-1}=200$ MeV
at temperature $37.5$ MeV. \ The results are in rather good agreement. \ In
the future we hope to use larger zone sizes and do simulations at lattice
spacings up to $a^{-1}=300$ MeV.%

\section{Summary and future directions}

We have introduced a new approach to the study of nuclear and neutron matter
which combines chiral effective field theory and lattice methods. \ Nucleons
and pions are treated on the lattice as point particles and we are able to
probe larger volumes, lower temperatures, and greater nuclear densities than
in lattice QCD. \ The low energy interactions of these particles are governed
by chiral effective field theory and operator coefficients are determined by
fitting to nucleon scattering data. \ Any dependence on the lattice spacing
can be absorbed by the renormalization of operator coefficients. \ In this way
we have a realistic simulation of many-body nuclear phenomena with no free
parameters, a systematic expansion, and a clear theoretical connection to QCD.
\ We have presented results for the energy per neutron for hot neutron matter
at temperatures 20 to 40 MeV and densities below twice nuclear matter density.

In conjunction with other members of the Nuclear Lattice Collaboration, we
plan several extensions, generalizations, and improvements upon this work.
\ In the course of producing data for this article, we have also generated a
large amount of data for the neutral pion, neutron, and pair density
correlation functions. \ This data will be analyzed and presented in a
forthcoming article. In the near future we also plan to study the pionless
version of the same neutron system. \ There has been a recent mean-field
discussion of this model \cite{Chen:2003vy}. \ Without pions, the phase will
be completely eliminated from the matrix determinant. \ This is due to the
fact that the matrix is purely real, and in the Hubbard-Stratonovich formalism
the up and down spins appear\ in a way that the matrix determinant is the
square of a real number. \ We also plan to extend our studies to include
neutrons, protons, neutral and charged pions, and make use the recent progress
in implementing exact non-linear representations of chiral symmetry on the
lattice \cite{Chandrasekharan:2003wy}\cite{Borasoy:2003pg}\cite{Lewis:2000cc}%
\cite{Shushpanov:1998ms}.

\begin{acknowledgments}
The authors benefitted from many discussions with members of the Nuclear
Lattice Collaboration and the organizers and participants of the INT Workshop
on Theories of Nuclear Forces and Nuclear Systems, Fall 2003. \ We especially
thank Boris Gelman, Ryoichi Seki, Robert Timmermans, and Bira van Kolck. This
work was supported in part by DOE grant DE-FG-88ER40388, NSF grant
DMS-0209931, and the Deutsche Forschungsgemeinschaft.
\end{acknowledgments}

\bibliographystyle{h-physrev3}
\bibliography{NuclearMatter}

\end{document}